\documentclass[usenatbib]{mn2e}

\usepackage{epsfig}
\usepackage{amsmath}
\usepackage[pdftex,bookmarks=true]{hyperref}
\newcommand{\mean}[1]{\overline{#1} }

\newcommand{\Cseed}{ C_{\rm{seed}}}

\newcommand{\Cmerge}{ C_{\rm{merge}}}

\begin{document}

\title[De-Trending Time Series for Astronomical Variability Surveys]{De-Trending Time Series for Astronomical Variability Surveys}

\author[Kim et al.]
{\parbox{17.5cm}{Dae-Won Kim$^{1, 2, 3}$\thanks{\href{mailto:dakim@cfa.harvad.edu}{dakim@cfa.harvard.edu}},
Pavlos Protopapas$^{1, 2}$, Charles Alcock$^{1}$, Yong-Ik Byun$^{3}$, Federica B. Bianco$^{1,4}$}\vspace{0.5cm} \\
$^{1}$Harvard Smithsonian Center for Astrophysics, Cambridge, MA, USA \\
$^{2}$Initiative in Innovative Computing, Harvard University, Cambridge, MA, USA \\
$^{3}$Department of Astronomy, Yonsei University, Seoul, Korea \\
$^{4}$Department of Physics and Astronomy, University of Pennsylvania, Philadelphia, PA, USA}

\date{Accepted ????; Received ????}

\pagerange{\pageref{firstpage}--\pageref{lastpage}} \pubyear{2008}

\maketitle
\label{firstpage}

\begin{abstract}
  \normalsize{ We present a de-trending algorithm for the removal of trends in time series.
  Trends in time series could be caused by various systematic and random noise sources such as cloud passages,
  changes of airmass, telescope vibration, CCD noise or defects of photometry.
  Those trends undermine the intrinsic signals of stars and should be removed.
  We determine the trends from subsets of stars that are highly correlated among  themselves.
   These subsets are selected based on a hierarchical tree  clustering algorithm.
  A bottom-up merging algorithm based on the departure from normal distribution in the correlation is developed to identify subsets, which we call clusters.
   After identification of clusters, we determine a trend per cluster by weighted sum of normalized light-curves.
  We then use quadratic programming to de-trend all individual light-curves based on these determined trends.
  Experimental results with synthetic light-curves containing artificial trends and events are presented.
  Results from other de-trending methods are also compared.
  The developed algorithm can be applied to time series for trend removal in both   narrow and wide field astronomy.}
\end{abstract}


\section{INTRODUCTION}

 Small-aperture telescopes have detected a large number of   exo-planet transits
 \citep{Alonso2004,Bakos2004,McCullough2005,Bakos2007,Burke2008,Pal2008,Pollacco2008}.
 A large number of variable stars have also been detected by surveys that use  such telescopes
 \citep{Schmidt1991,Akerlof2000,Pojmanski2005,Schmidt2007,Pigulski2008,Szczygiel2008}.
  A weakness in these surveys is that the signal to noise ratio (S/N) is lower than   the S/N obtained by larger-aperture telescopes.
  The low S/N can be  attributed not only to the small-aperture size
  but also to noise in CCD images such as non-uniform illumination, or to local weather
  changes throughout the field (especially in the case of wide field surveys).
  To improve the S/N and thus improve the detectability of variability,   these noise sources should be minimized.

Some of these  noise sources are strongly correlated between light-curves of different stars.
  For example, if a star appears fainter, other stars near it may appear fainter at the same time.
  We call such coherent changes through parts of the field {\em trends}.
  These trends could be caused by local weather patterns
  such as thin cloud passages or airmass changes \citep{Howell1986, Kjeldsen1992, Gilliland1988} throughout the night.
  The conventional approach for trend removal is differential photometry with a {\em reasonable} selection of template stars near the star of interest   \citep{Young1991, Everett2001}.
  With the help of modern CCDs, it is not   hard to select a sufficient number of bright stars as a template set.
  However, the de-trended results are then sensitive to the selection of template stars.
  If the template stars contain intrinsic variables, the determined trends will be different from  the true trends.
  Therefore, excluding such intrinsically variable stars from template stars is essential.
  Furthermore, because there is no guarantee that trends are the same for all stars throughout the entire field,  the template selection method should be able to handle localization of trends in large fields of wide field surveys.

  In this paper, we propose a new de-trending method   (hereafter PDT, for  Photometric De-Trending algorithm),
   which incorporates a systematic template   selection algorithm that can solve the problems mentioned above and consequently shows superior de-trended results.
  Experiments with simulated light-curves show that PDT correctly reproduces localization.

  We present details of PDT in Sec. \ref{sec:Algorithm}.
  In Sec. \ref{sec:Test}, we show de-trended results for synthetic   light-curves containing artificially added trends and events.
  In addition, comparison results with the Trend Filtering Algorithm (hereafter, TFA) \citep{Kovacs2005} are  also presented.
In Sec. \ref{sec:TAOSandMMT}, we show two examples of astronomical datasets and their de-trended results.
We outline future work in Sec. \ref{sec:NoteFutureWork}.  We summarize our  conclusions  in Sec. \ref{sec:Conclusion}.

\section{Algorithm}
\label{sec:Algorithm}
\subsection{Outline of the PDT}

 One of the most widely used methods for the selection of template stars is the method that chooses as
 a template set a sufficient number of bright stars that are not saturated, not overlapping  and  not at the edge of the field.
 Some of these bright stars could have  intrinsic variability (e.g. variable or flare stars).
 If we avoid those stars in the selection of template set, the de-trended results will be improved.
 Ideally, standard stars such as Landolt standard stars \citep{Landolt1992} could be useful as template set. However, there are not many standard stars in the field (even in the wide field surveys). Thus, one needs to choose template stars from the field,
 where a few \% of stars are varying (see for example \citealt{Paczynski2000, Everett2002}).

 If the light-curve of a star manifests a trend without being intrinsically variable,  then the light-curve should be
 highly correlated with many other light-curves of stars in that field.
If a star has both trend and intrinsic variability,  the light-curve of the star would not be as highly correlated with other light-curves.
 Therefore, a light-curve which has strong correlation with many other light-curves is a good template candidate.
 Our approach to the selection of template stars  is to choose highly correlated subsets of stars using the similarity matrix $C$,
 in which  the elements $C_{ij}$ are the Pearson correlation values between light-curves of star $i$ and star $j$.

 The Pearson correlation values can be calculated by the following equation:

\begin{eqnarray}
\label{eq:CM} C_{ij} = \frac{1}{n-1} \,\, \frac{
\displaystyle\sum_{t=1}^{n} ( L_{i}(t) L_{j}(t)) - n   \mean{L}_{i} \,
\mean{L}_j }  {\displaystyle\sigma_i \sigma_j},
\end{eqnarray}

{\noindent}where $L_i(t)$ is the flux  of star $i$ at time $t$, $n$
is the total number of measurements, $ \mean{L}_i$ is the mean flux of
$L_i(t)$ and $\sigma_i$ is the standard deviation of $L_i(t)$. The
number of measurements $n$ for every light-curve should be the same.

Using the similarity matrix and a hierarchical tree clustering
algorithm explained below, we can extract multiple subsets of template stars;
each subset is relatively highly correlated within itself but not with any other subsets. We call the subsets {\em clusters}.
For each extracted cluster, we determine one representative trend light-curve
by the weighted sum of all  light-curves from that cluster.
 To remove the trends from all light-curves, we minimize the residuals between
 each light-curve  and the determined trends  by minimizing the root mean square (rms) $r_i$,

 \begin{eqnarray}
 \label{eq:rmsmin}
  r_i = \sqrt{\frac{1}{n} \sum_t \left[ L_i(t) - \lambda_i - \sum_{k}^m \beta_{ik} \, T_k(t) \right]^2},
\end{eqnarray}

{\noindent}where $n$ is the total number of measurements,
$T_k(t)$ are the determined trends for cluster $k$, $m$  is the
total number of clusters, $\beta_{ik}$ and $\lambda_i$ are free
parameters to be calculated for each light-curve. For more details
about the minimization process, see Sec.
\ref{sec:DetermineRemovalTrend}.

Sometimes  such minimization
approaches remove not only trends, but also the intrinsic signals
because one can adjust the free parameters such that the summed trends
resemble the signals.
 This side effect is more significant when there are more free parameters to be adjusted.
 Therefore, PDT, which identifies one representative trend per cluster and thus has a small number of free parameters,
 is better suited for de-trending light-curves,   especially where the rms contribution from the  intrinsic signal is significant.
 This contrasts with TFA or similar methods that assign   one free parameter per template star per individual light-curve.

 In the following sections, we explain how we use the similarity matrix to choose the clusters and
 how we de-trend light-curves using the selected clusters.

\subsection{Selection of Clusters of Light-Curves}
\label{sec:Clusters}

First, we summarize  traditional clustering algorithms and their
shortcomings in Sec. \ref{sec:TraditionalClustering}. We then
explain a selection method for choosing clusters of light-curves
using a hierarchical tree clustering algorithm, which is more
suitable  than other clustering algorithms. The selection method
consists of two processes. The first step is the construction of a
hierarchical tree according to the similarity matrix, explained in detail in Sec.
\ref{sec:HierarchicalClustering}. The second step is the extraction
of clusters from the constructed hierarchical tree using the
normality test explained in Sec. \ref{sec:AgglomerativeAlgorithm}.

\subsubsection{Clustering Algorithms}
\label{sec:TraditionalClustering}

 In order to extract clusters of template stars, we first group stars using a clustering algorithm based on the similarity matrix.
 Clustering algorithms are useful for grouping large data according to their similarities \citep{Jain1999}.

 We have examined several clustering algorithms, such as density-based clustering \citep{Ester1996},
 K-mean \citep{Hartigan1979}, K-medoids (also known as Partitioning around Medoids or CLARANS, \citealt{Ng1994})
 (hereafter K-methods) and a hierarchical tree clustering algorithm \citep{Jain1999}.
 These algorithms first define distances between each element (light-curves  in our case)
 and then group elements that are similar to each other based on the distance.
  For all our testing, we used a distance matrix in which the elements are defined as:

 \begin{eqnarray}
  \label{eq:Distance}
 D_{ij} \equiv 1 - C_{ij},
 \end{eqnarray}

 {\noindent}where $C_{ij}$ are the Pearson correlation values between  two elements $i$ and $j$,
 as shown in equation \ref{eq:CM}.
 More correlated, or more similar elements have shorter distances between them.

\begin{figure}
\begin{center}
        \includegraphics[width=0.4\textwidth]{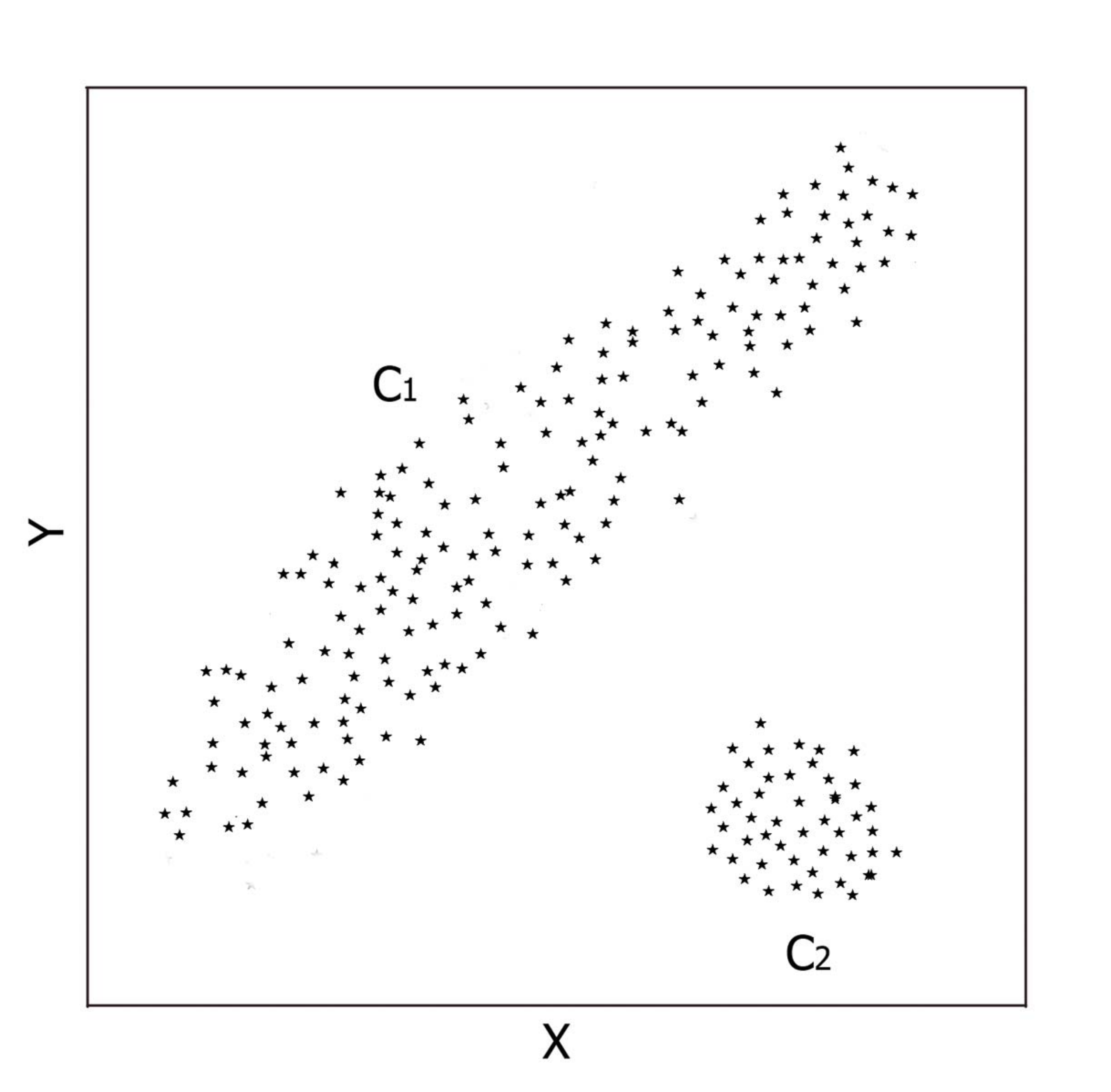}
\end{center}
    \caption
           {Conceptual illustration of the problem with most clustering methods applied to de-trending.
           Using  most algorithms, two clusters in the figure can be easily identified.
           Even though some of the elements in cluster $C_1$ are far  from each other, they are identified as one cluster.
           The x and y-axes indicate the distances between pairs of elements.}
           \label{fig:DBC}
\end{figure}

 In choosing template sets, it is  important while grouping elements that every
 element in the same cluster is similar to the others in the cluster.
 However,  some of the clustering algorithms \citep{Hartigan1979, Ester1996, Ng1994}
 group into clusters elements which are not pairwise similar.
 This is a critical disadvantage because we would like to identify only stars that are strongly correlated to one another.
 Fig.  \ref{fig:DBC} conceptually illustrates  the problem.
 The x and y-axes indicate the distances between pairs of elements, where closer elements are more similar.
 By means of these clustering algorithms,  one can easily identify the two clusters, $C_1$ and $C_2$, in Fig. \ref{fig:DBC}.
 Yet, some elements in the cluster $C_1$ are not close to other elements in the same cluster
 because the cluster $C_1$ is stretched along the diagonal direction.
 For example,  the bottom left elements are far from the top right elements, even though they are in the same cluster.
With the exception of the hierarchical tree clustering algorithm,
the clustering methods mentioned above suffer from these disadvantages.

Note that the term `cluster' in this paper is not used in the conventional way, where  $C_1$ in Fig. \ref{fig:DBC} would be considered as a cluster.
In the rest of paper, we will be using the term `cluster' to designate `zone of influence' which means a group of strongly correlated elements.
In this concept, $C_1$ would be split into several smaller sub-groups.

\subsubsection{Hierarchical Tree Clustering Algorithm}
\label{sec:HierarchicalClustering}

 A hierarchical clustering algorithm is substantially different from density-based clustering or K-methods.
 It constructs a hierarchical tree by linking all elements  together under the same root according to predefined distances (see equation \ref{eq:Distance}).
 During the construction, it does not need to estimate initial parameters such as
 the minimum number of elements (as in density-based clustering algorithms)
 or the total number of clusters (as in K-methods). This is an advantage of the hierarchical algorithm.

\begin{figure}
\begin{center}
        \includegraphics[width=0.48\textwidth]{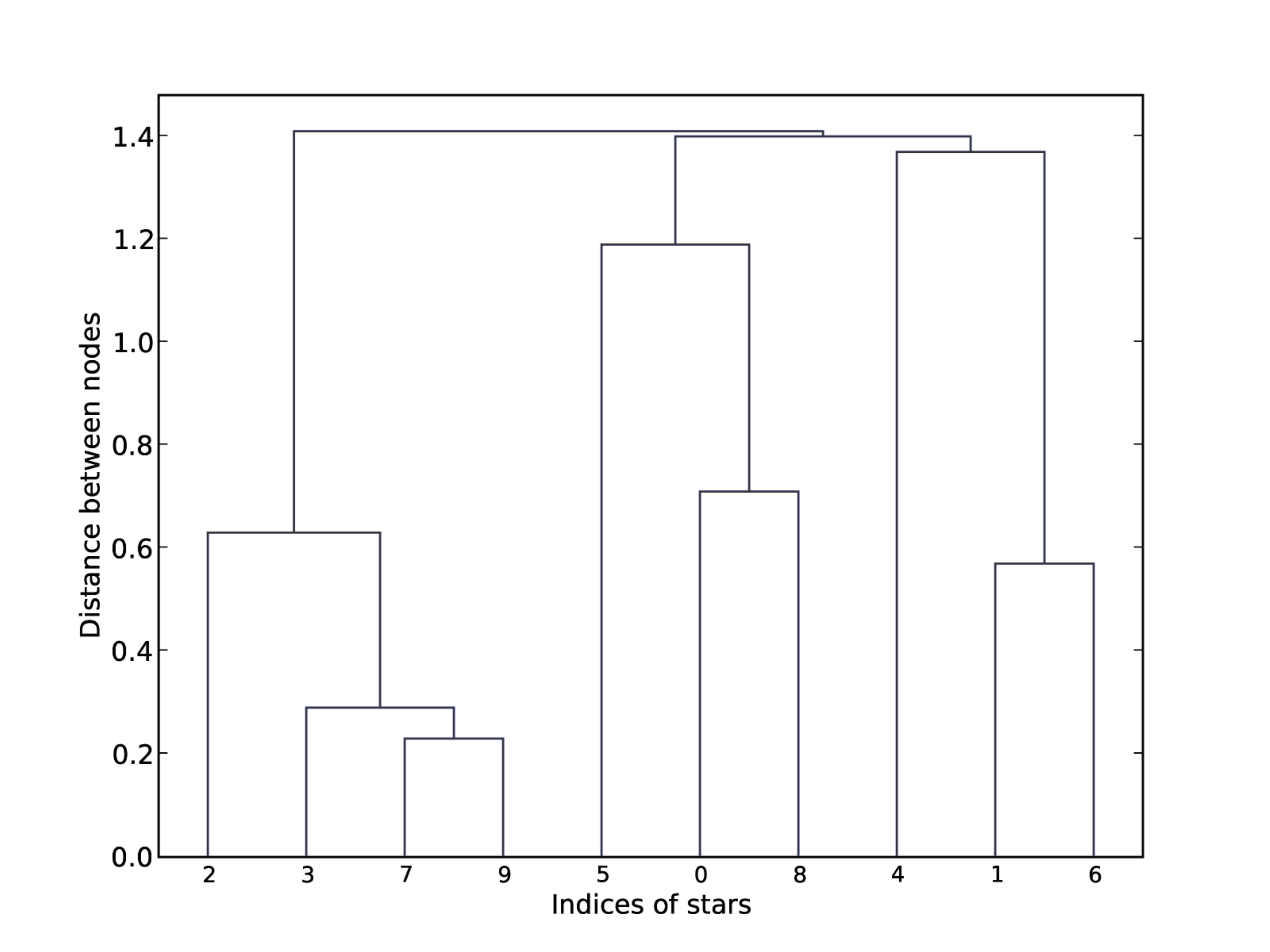}
\end{center}
    \caption
           {An example of a dendrogram.
          The x-axis is the index of the star, and the y-axis is the distance between nodes.}
    \label{fig:Dedrogram}
\end{figure}

 The constructed hierarchical tree is traditionally represented by a dendrogram as shown in Fig. \ref{fig:Dedrogram}.
 We use the predefined distance matrix in order to link elements and generate the dendrogram.
 At each stage of linkage, the algorithm joins the two closest nodes into a new set.
 The `node' can consist of either a single element or previously connected  multiple elements.
 This process continues until all elements belong under the same root.
 During this linkage process, we need to define the distances between two nodes as well.
 There exist several methods to calculate the distance between nodes \citep{Jain1999}.
 Among these methods, we use the complete-linkage method to construct the tree.
 In the complete-linkage method, the distance  between two nodes is defined as the longest distance
 among the pairwise  distances between the elements (as defined in equation \ref{eq:Distance}) of the two nodes.
 Therefore, the distance between any two elements in two nodes is always smaller or equal to the distance between two nodes.
 The complete-linkage method was chosen  because it produces more tightly bound clusters
 and hierarchies  than  other methods such as the single-linkage method \citep{Jain1999}.

 Fig. \ref{fig:Dedrogram} shows an example of a dendrogram of a hierarchical tree constructed by the complete-linkage method.
 We plot only 10 elements in Fig. \ref{fig:Dedrogram} as an example.
The  x-axis is the index of each star and the y-axis is the distance between
nodes. The height of the horizontal lines in the dendrogram
represents the distance between two nodes linked together.
 We used the
 \href{http://bonsai.ims.u-tokyo.ac.jp/~mdehoon/software/cluster/software.htm\#pycluster}{\fontfamily{pcr}\selectfont PyCluster} library \citep{deHoon2004}
 to generate the hierarchical tree and the \href{http://code.google.com/p/scipy-cluster/}{\fontfamily{pcr}\selectfont hcluster} library to draw the dendrogram.

 Traditionally, hierarchical algorithms do not produce clusters, unlike  the other clustering algorithms that group elements into resulting clusters.
 This is a conventional feature of hierarchical algorithms \citep{Daniels2006}  and it means that users must decide which elements in the tree should be grouped into resulting clusters.
This is equivalent to defining the number of clusters in K-means clustering  or defining the  connectable distance in the  density based clustering.
To solve this problem, we propose an extension to the  hierarchical algorithm, shown in the following section, that can extract the resulting clusters  from the tree without the need of predefining such parameters.

\subsubsection{Agglomerative Merging Algorithm for Selection of Clusters}
\label{sec:AgglomerativeAlgorithm}

 With the constructed dendrogram in hand, we can link every star according to the distance matrix.
 We now need to extract subsets of stars that are  highly correlated among  themselves for
 a template set and to exclude outliers such as intrinsic variables  that can be harmful for de-trending.
 Furthermore, if there exist multiple and different trends in data,  we should be able to separate them as well.
The traditional method to achieve this is to set a certain distance value
and extract subsets such that the farthest distance between
elements in the subset is smaller than the set  distance
(e.g. subset [3, 7, 9] will be extracted given a set  distance = 0.4 in Fig. \ref{fig:Dedrogram}).
On the other hand, it is not easy to choose a
set distance, especially for different datasets, for example with  data
observed under different weather conditions,
different dates, or with different telescopes.
 As we mentioned in previous section, this is a conventional feature of the hierarchical tree algorithm in extracting relevant
 and representative clusters.

To alleviate this problem, we developed an agglomerative merging
algorithm (bottom-up merging algorithm)  to identify the clusters in
the constructed tree, based on the assumption of normal distribution \citep{Kim2008}.

 First, we note that distances between correlated light-curves follow a skewed distribution
 in contrast to the distribution of distances of uncorrelated light-curves  that is known to follow a normal distribution.

 Second if one applies Fisher's transformation \citep{Fisher1915},

  \begin{eqnarray}
 C'_{ij} = \displaystyle \frac{1}{2} \log \frac{1 + C_{ij}}{1 - C_{ij}} \, \, ,
 \label{eq:fisher}
 \end{eqnarray}

{\noindent}to the correlation values $C_{ij}$,  the resulting transformed $C'$s are approximately normally distributed
\citep{Anderson1996}.

 Now we can claim that if a single cluster comprises correlated light-curves and does not contain outliers,
 the transformed distances between the light-curves in the cluster are normally distributed.
 We then extract subsets by merging the two closest nodes  that have the shortest distance in the tree (see details of the process below).
 We repeat the merging processes and test the normality at every merging step to decide whether  to stop the merging processes.
 To test  normality, we use the Anderson-Darling test \citep{Anderson1952,Stephens1974} which tests the null hypothesis
 that a dataset comes from the normal distribution. In other words, the test can statistically quantify
 how far the dataset departs from the normal distribution.
 Based on the test, one can derive the p-value
 that indicates the level of significance of the departures from normal distribution \citep{D'Agostino1993}.
 If the subset fails the normality test,  it is inferred that there exist outliers in the subset
 or the subset consists of two or more different trends.
 Therefore, we stop the merging process  below the  level where the normality test fails.
 If we repeat this process for extracting subsets in the hierarchical tree,
 we can finally obtain multiple clusters of trends without outliers.

 Realistically, there is a mixture of various noise sources including Poisson noise and trends,
 and thus the distances between light-curves in a cluster
 might not be perfectly normally distributed even after Fisher's transformation.
Also, because the correlation coefficients in a given cluster are not totally independent (e.g.
$C_{12}$ and $C_{13}$ are not totally independent of $C_{23}$),
and because we repeat the p-value testing on the same subset multiple times,
the p-value  should be considered as a tuning parameter (threshold) instead of its strict
statistical definition.
Nevertheless, using the normality test, we can extract strongly
correlated elements that are placed in the central part (peak) of the distribution.
Note that only strongly correlated elements are important to determine trends.

\vspace{0.2cm} We describe the details of the agglomerative merging algorithm here:
\begin{enumerate}
 \item Select initial cluster seeds to be all nodes which consist of only two elements in the constructed tree (e.g. [7, 9], [0, 8] and [1, 6] in Fig. \ref{fig:Dedrogram}).
\label{item:s2}

\item \label{item:seed} Define $\Cseed$  to be the node that has the shortest distance between two elements among selected cluster seeds from step \ref{item:s2}.

 \item \label{item:merge} Merge $\Cseed$ with its next linked node in the tree and call it $\Cmerge$.
 If the number of elements in $\Cmerge$ is smaller than 5, keep merging with the next linked node.
This is because if the number of elements in  $\Cmerge$ is too
small,  the normality test would not be reliable.

  \item \label{item:normal} Apply the Anderson-Darling test to the distance list of $\Cmerge$\ and derive the
  p-value.\footnote{We use   \href{http://www.r-project.org/}{\fontfamily{pcr}\selectfont R} statistical packages
  and \href{http://rpy.sourceforge.net/}{\fontfamily{pcr}\selectfont RPy} library to calculate the p-value.}
  The distance list is the list of all distances  between members of $\Cmerge$.
  For instance, if the indices of members in $\Cmerge$ are [1, 2, 3],   the distance list is [$D_{12}, D_{13}, D_{23}$] where   $D_{ij}$
  is the distance matrix we defined in equation \ref{eq:Distance}.
  We apply the Fisher's transformation before we apply the normality test as we mentioned above.

 \item If the calculated p-value is bigger than 0.1,  which means we cannot reject the null hypothesis that the distribution is normal
 with the significance level of 10\%,  set $\Cmerge$ as new $\Cseed$ and go to step \ref{item:merge}.
 Otherwise, stop the merging process and go to step \ref{item:save}.

 \item \label{item:save}Identify $\Cseed$ as cluster candidate.
 Go to step \ref{item:seed} and choose the next closest pair.
 Keep these processes until there remain no initial cluster seeds.

 \item \label{item:dup} Remove duplicated clusters from the candidates list derived at step \ref{item:save}.
 The duplication can happen when there exist multiple seeds in one cluster, that  can yield identical clusters.
 Note that as long as the initial cluster seeds defined in step \ref{item:s2} are the same,  the resulting clusters are the same no matter which cluster seed we start from.

 \item \label{item:nmin}  Remove clusters whose number of elements are smaller than 10.
 We need a sufficient number of elements (light-curves) to cancel out the uncorrelated noise in the light-curves  while determining master-trends (see Sec. \ref{sec:DetermineRemovalTrend}).

 \item  Define the list of clusters from step \ref{item:nmin} as $C_k$, where $k$ is the index of each cluster.

\end{enumerate}

The  clusters identified by the algorithm above are used to determine trends which we explain in the following section.

\vspace{0.3cm}
 While testing the merging algorithm, we observed that if we constrain the initial seeds at step \ref{item:s2},
 we can improve our algorithm by 1) decreasing CPU processing time and 2) removing  relatively contaminated clusters
 by other noise sources such as Poisson noise.
 We explain the details below.

  If we select the initial seeds to be just the pairs of elements whose distances are smaller than
   the average value of the distance matrix ($\mean{D}$) at step \ref{item:s2},
   we obtain a smaller number of seeds that are more highly correlated.
 $\mean{D}$ is given by:

 \begin{eqnarray}
 \label{eq:averagecutting}
 \mean{D} = \frac{1} {\displaystyle N(N-1) } {\displaystyle\sum_{i=1}^{N-1}\displaystyle\sum_{j=i+1}^{N}D_{ij}}  \,,
 \end{eqnarray}

 {\noindent}where $N$ is the total number of light-curves.
 The benefit is that we speed up the algorithm by reducing the number of iterative processes
  that  mainly consist of merging nodes and testing for normality.
 As we explained above, we repeat the merging and the normality test for every  initial seed.
  Therefore, if there are fewer initial seeds, there are fewer iterative processes,  thus reducing the CPU processing time.
 Moreover,  we can remove pairs of faint stars in advance from initial seeds.
 Faint stars suffer from noise more than bright stars, therefore,
 the clusters  derived with pairs of faint stars are less suited to determine trends
 than the clusters derived with pairs of bright stars.
  Note that if we use a looser constraint ($\gg\mean{D}$) and thus have too many seeds,
  the number of weakly correlated clusters and  the computational cost will increase.
 On the contrary, if we use a tighter constraint ($\ll\mean{D}$) and thus fewer initial seeds,
 we may  miss real clusters. We empirically found that any cutting values from $\mean{D} / {10}$ to $\mean{D}$ give reasonable results.
  Within this range, the overall characteristics of the determined trends using the resulting clusters  were almost identical.

In addition, it is known that the square root of the variance of correlation coefficients are generally:
\begin{eqnarray}
\label{eq:stdCorr}
\sigma = \frac{1 - C_{ij}^2} {\sqrt n} \, ,
\end{eqnarray}
{\noindent}where $C_{ij}$ is the correlation value
between two variables and $n$ is the total number of measurements \citep{Bowley1928, Hotelling1953, Ghosh1966}.
If the light-curves consist of random fluctuations (e.g. pure Poisson noise), $C_{ij} \simeq 0$.
Thus, equation \ref{eq:stdCorr} changes to:
\begin{eqnarray}
\sigma \simeq \frac{1} {\sqrt n}
\end{eqnarray}

{\noindent}We remove all initial seeds from step \ref{item:s2} whose distances are larger than  $1 - 3 * \sigma$  because resulting clusters using these initial seeds would  contain  light-curves of mainly random fluctuation that are not correlated with other light-curves. Note that this criterion is different from the one above.  For example, this occurs when  there is a set of light-curves of random fluctuations. In that case, $\mean{D}$ is $\sim$1 and several initial seeds whose distances are smaller than 1 would  pass the $\mean{D}$ criteria.

We also tested another threshold cut which constrains
elements in each cluster to be highly correlated.
If a distance between any two elements in a given subset
is bigger than $\mean{D}$, we stopped the merging process
even if the subset was not rejected by the normality test.
Nevertheless, we empirically found that resulting clusters
and de-trended light-curves are not affected by this threshold.

\subsection{Determination and Removal of Trends}
 \label{sec:DetermineRemovalTrend}

 With the extracted single or multiple clusters,  we next determine the trends for each cluster (hereafter, master-trends),
 from the weighted sum of the cluster members as:

\begin{eqnarray}
\begin{array}{rcl}
T_{k}(t) & = & \frac{\displaystyle\sum_{i=1}^{N_{k}} \, w_i\,
f_i(t)} {\displaystyle\sum_{i=1}^{N_{k}} \, w_i},
 \\ \\
  f_i(t) & = &  \displaystyle \frac{{L_i(t) -   \mean{L}_i }} {\mean{L}_i}  \,\, ,  \\ \\
 w_i & = & \displaystyle\frac{1}{\sigma_{f_i}^2} \,\, , \\ \\
 \end{array}
\end{eqnarray}

{\noindent}where $\sigma_{f_i}$ is the standard deviation  of $f_i$,
$N_{k}$ is the total number of template stars in the cluster $C_k$,
 $t$ is the time index with the total $n$ measurements,
 $L_i(t)$ is the light-curve of $i^{th}$ template star and  $\mean{L}_i$ is the mean value of $L_i(t)$.

This master-trend set, $T_k(t)$, is used to de-trend the individual light-curves.
Each master-trend well represents the characteristic of each cluster
because all the light-curves in each cluster are selected to be strongly correlated.
Note that we determine just one master light-curve per cluster.

 After we determine the master-trends,  we remove the trends from each individual light-curve.
  First we normalize each light-curve $L_i(t)$ as:

 \begin{eqnarray}
 \hat{L}_i(t) =  \displaystyle \frac{{L_i(t) -   \mean{L}_i }} {\mean{L}_i} \,\, .
 \end{eqnarray}

We then assume that each light-curve $\hat{L}_i(t)$,  is a linear combination of
 the determined master-trends $T_k(t)$, and noise, $\epsilon_i(t)$,

\begin{eqnarray}
       \hat{L}_i(t)   =   \displaystyle\sum_{k=1}^{m}  \beta_{ik} \, T_k(t)  +\epsilon_i(t)   \,\, ,
\end{eqnarray}

 {\noindent}where $i$ are the indices of individual light-curves to be de-trended, $k$ are the indices of master-trends,
 $m$ is the total number of master-trends and $\beta_{ik}$ are free parameters to be determined
 by means of minimization of $\sum_t \epsilon_i(t)^2$ (equivalent to minimizing   $r_i^2$  in equation \ref{eq:rmsmin}).

 During the minimization of $\sum_t \epsilon_i(t)^2$, there is one more complication we have to consider.
 Let us assume there exists a single trend where flux increases monotonically
 and an intrinsic variable star where flux decreases monotonically.
 Even though the direction of the trend is different from that of the variable star,
 the minimization method will eventually reduce the intrinsic signal
 because the free parameters can take negative values and thus minimize  $\sum_t \epsilon_i(t)^2$.
 To eliminate this undesirable effect, we constraint the free parameters $\beta_{ik}$, to be always bigger than or equal to zero
 using quadratic programming \citep{Goldfarb1983}.\footnote{Quadratic programming is
 a mathematical optimization method which minimizes (or maximizes)
 a quadratic function of several unknown parameters which is subordinate to
 linear constraints on the parameters.  We use
 \href{http://www.r-project.org/}{\fontfamily{pcr}\selectfont R} statistical packages
 to implement quadratic programming.}

\section{Test with Synthetic Light-Curves}
\label{sec:Test}

 We present here the results from several simulations we performed.
  First, we describe the method by which we parameterized trends and how we built the simulation
  (see Sec. \ref{sec:DataDescription}). Next, we present de-trended results of artificially inserted  transits and eclipsing binaries
  using PDT (Sec. \ref{sec:simulIdenCluster}) and  comparison results with TFA  (Sec. \ref{sec:transit}).
  Finally, we show simulations and results from other unique configurations (Sec. \ref{sec:secondANDotherconsideration}).

\subsection{Data Description}
\label{sec:DataDescription}

 We generated $\sim$500 artificial light-curves, each  having different flux and 360 one-minute-exposures.
 During this simulation, we set x-coordinates of stars as altitude and y-coordinates as azimuth.
 CCD size was set to 2048x2048.  The magnitudes of the stars were chosen   from the USNO B1.0 catalog   \citep{Monet2003}
 within a particular  patch of the sky   ($3 \deg^2$) [$4^h \, 48^m \, 00^s, \, 20^\circ \, 46' \, 20''$]
 and ranged from $\sim$6 mag to $\sim$13 mag.
Poisson noise was added to light-curves with  standard deviation
values ($\sigma$) set to vary from  0.001 mag to 0.02 mag.
Although there exist other possible sources contributing to the noise budget such as CCD overscan,
bias, dark, flat-field, etc. \citep{Gilliland1988},
we did not include those noise sources because trends are predominantly
due to weather (sky background).  Bias and other such error sources
are usually stable during a night observation and not a major source of trends.
The results of this simulation would not be affected by such noise, as can be seen
from the analysis presented in Sec. \ref{sec:TAOSandMMT}, where we used non-simulated data that contain bias, etc.

 We added three transit signals \citep{Mandel2002} into three different light-curves with $\sigma=0.01$ mag.
 Transit depths were 0.015, 0.020 and 0.025 mag  with 60 minutes duration (one-sixth of total observation duration).
 We placed the transit signals at the central part of the light-curve.
 We also added two eclipsing binaries into two light-curves with  $\sigma = 0.01$ mag.
The remaining stars were set to  have no intrinsic variability.

\vspace{.4cm} \noindent To add trends, we artificially generated four types of trends:

\begin{enumerate}
\item \noindent  \label{item:firsttrend} \underline{first order atmospheric extinction.}
This is the typical extinction that linearly depends on the airmass:
$a \, M_i(t),$ where $a$ is the extinction coefficient, which is
$\sim$0.16 for the V and $\sim$0.1 for the R band \citep{Stalin2008},
 $M_i(t)$ is the airmass  of $i^{\rm th}$ star and is given by:
\begin{eqnarray}
\begin{array}{rcl}
M_i(t) & = & \sec(z_i(t)) \,\, ,\\ \\
z_i(t) & = & 90^\circ - [ (c + d \, t) + e \, \hat{y}_i] \,\, ,
\end{array}
\end{eqnarray}

\noindent where $c$ is the starting altitude of the field
($c=45^\circ$), $d$ is  the change of altitude per minute
($d=0.25^\circ \min^{-1}$), e is the field of view ($e=3\deg^2$),
$\hat{y}_i = y_i / D_y$ is the y-position of $i^{th}$ star
normalized by y-size $D_y$ of CCD plane and $t$ is the observational
time in minutes.

To change airmass with time, we changed the altitude of the field
from $-45^\circ$, passing through $90^\circ$, to $45^\circ$.

\vspace{0.2cm}

\item \noindent \label{item:secondtrend} \underline{position-dependent and time-dependent trend.}
We model this type of extinction to imitate  stationary `clouds'
and thus to depend on the azimuthal position of the star and
observation time as: $b \, \hat{t} \, \hat{x}_i  ,$ where $b$ is the
maximum depth (for this simulation we use $b=0.01$ mag),
$\hat{t}=t/t_{\rm{TOTAL}}$, is the normalized time over the total
observation duration ($t_{\rm{TOTAL}}$),
 $\hat{x}_i = x_i/D_x$ is the x-position
of $i^{\rm{th}}$ star normalized by $D_x$, the x-size of CCD plane.
Such position-dependent and time-dependent  trends can be caused by
thin cloud, Moon light or occasionally by CCD noise.

\vspace{0.2cm}
\item \noindent \label{item:thirdtrend} \underline{localized trend.} This is an artificially CCD-localized trend that has a simple
linear time dependance, $\zeta_i(t)$, given by:
\begin{eqnarray}
\zeta_i(t)  &=& f \hat{t}\, ,\,  \\ \nonumber
 f   &=&
  \left\{ \begin{array}{ccccc}
    0.25  &  & \,\, x_i > 1500 & \,\, \rm{ and } \,\,  & y_i < 500
    \\
    0 & & \rm{otherwise} & & \\
 \end{array} \right.
\end{eqnarray}

{\noindent}where $x_i$ is the x-position of $i^{th}$ star, $y_i$ is the y-position of $i^{th}$ star and $\hat{t}$ is the normalized time as
explained above. Such localized trends can be caused by non-uniform
clouds or the non-uniform illumination structure of CCD images.

\hspace{0.2cm}
\item \noindent  \label{item:fourthtrend} \underline{the second order atmospheric extinction.}
This is the other atmospheric extinction related to the star color,
$w r  C M_i(t)$, where $w$ is proportional to the square     of the
optical bandwidth, C is a color index and $r$ is the difference between
the extinction coefficients in the corresponding bands
\citep{Young1991}. The coefficients $w$ and $r$ are constant and
same for all stars in the field.
    Even if the airmass changes for two stars are same during observation (e.g. two stars at same altitude),
    trends could be different due to the differences in colors
    (typically a few milli-magnitude differences in light-curves, \citealt{Young1991}).
 We will ignore this term until Sec. \ref{sec:secondorderextinction}.

\end{enumerate}

\begin{figure}
\begin{center}
       \includegraphics[width=0.45\textwidth]{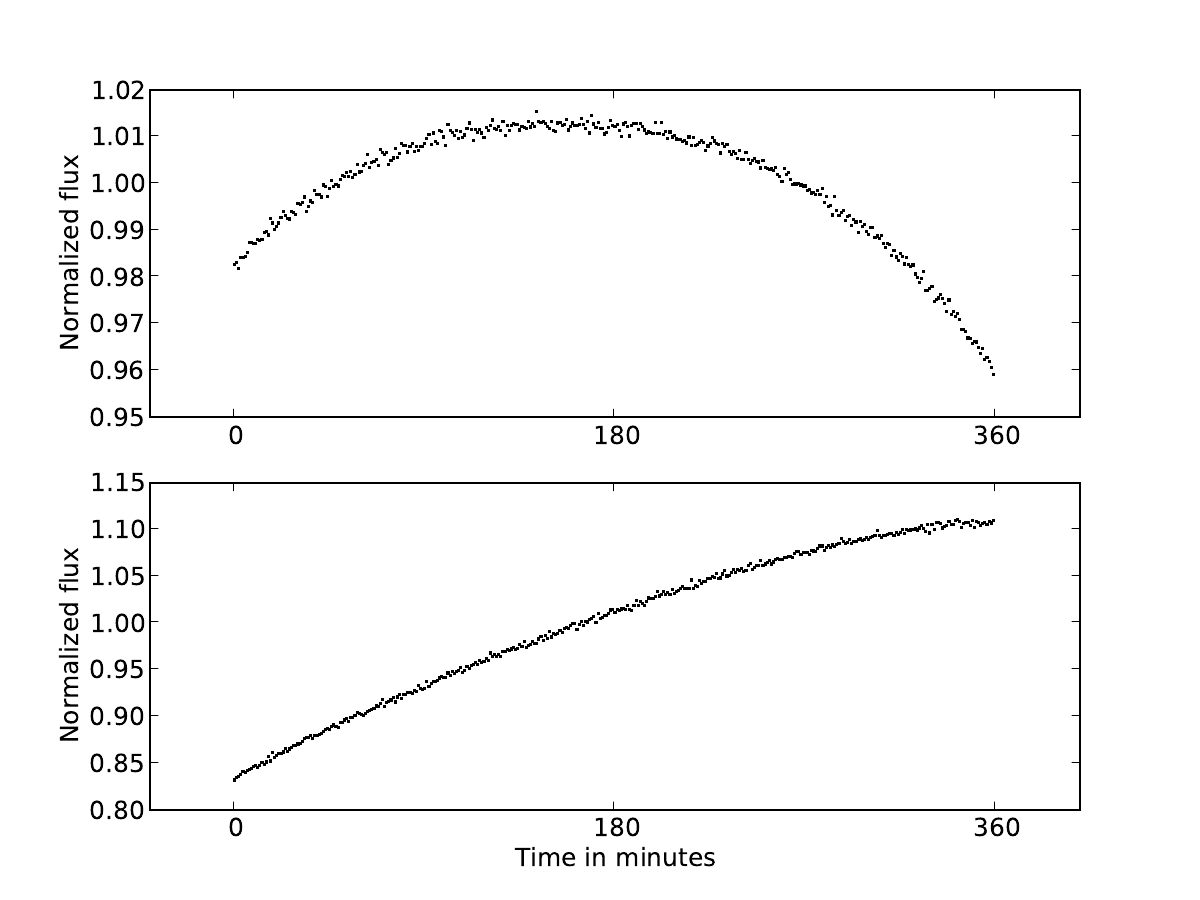}
\end{center}
    \caption{Two sample synthetic light-curves of two bright stars which contain trends.
    The light-curve at the top panel consists of the first order atmospheric extinction,
    position-dependent and time-dependent trend, and Poisson noise.
    The light-curve at the bottom panel contains an additional trend that is artificially CCD-localized.}
    \label{fig:trendSample}
\end{figure}

 Fig. \ref{fig:trendSample} shows two distinctive trends build on two  bright stars.
The top panel is a light-curve of a bright star which consists of the
first trend, the second trend (\ref{item:firsttrend},
\ref{item:secondtrend}) and Poisson noise. The bottom panel is a
light-curve of another bright star which consists of  the first trend,
the second trend, the third trend (\ref{item:firsttrend},
\ref{item:secondtrend}, \ref{item:thirdtrend}) and  Poisson noise.

\subsection{Identification of Clusters of Template Stars}
\label{sec:simulIdenCluster}

\begin{figure}
\begin{center}
       \includegraphics[width=0.45\textwidth]{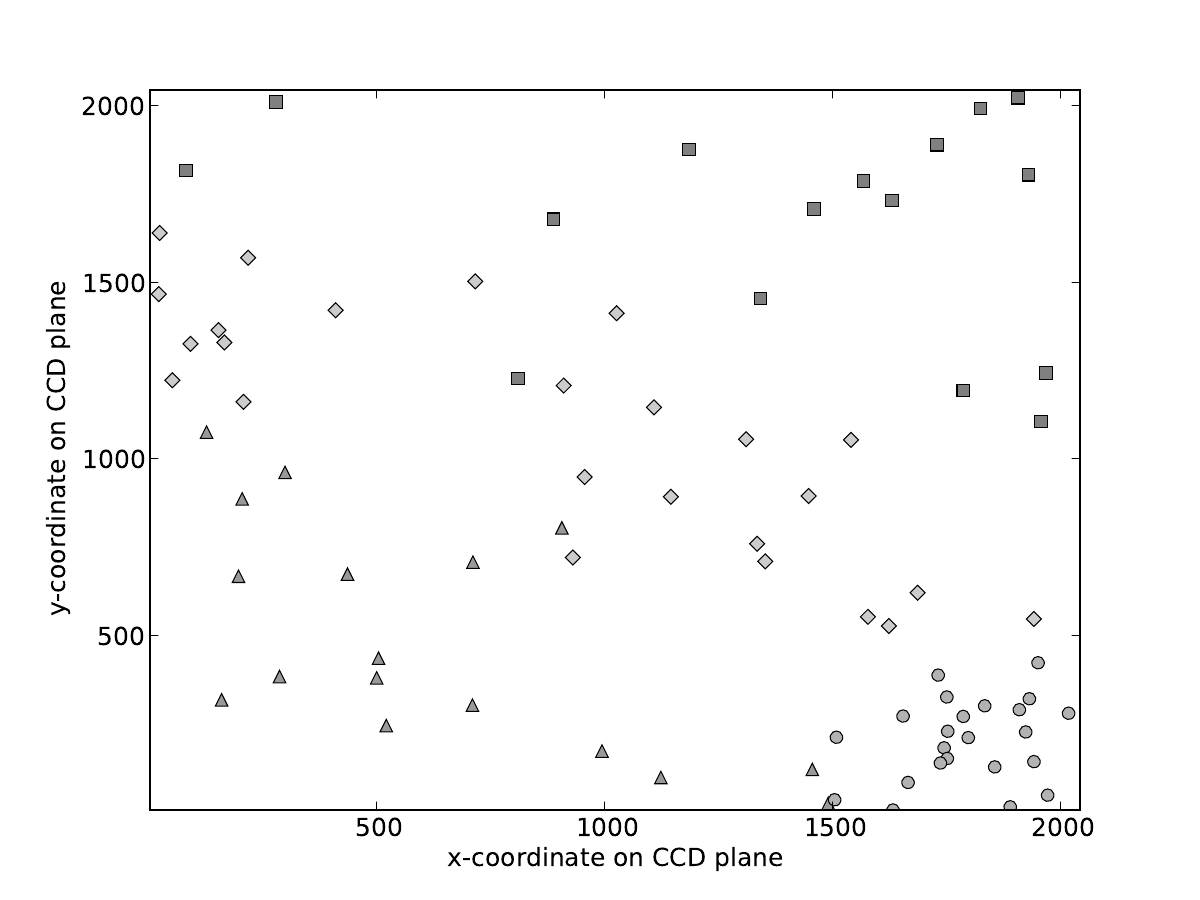}
\end{center}
    \caption
           {
            Positions of the identified four clusters in the artificially generated dataset. x(y)-axis is the x(y)-coordinates of stars on the
            field. Four different shapes mean four different clusters.
           }
    \label{fig:TP}
\end{figure}

 We applied PDT to test its ability to properly identify the inserted artificial trends.
 Using PDT, we identified four different clusters in the dataset as shown in Fig. \ref{fig:TP}.
 The x and y-axis of Fig. \ref{fig:TP} are the x and y-coordinates of the template stars on   the CCD plane.
 Different symbols indicate different clusters. Each cluster is well separated along the y-axis due to  the artificially inserted airmass (\ref{item:firsttrend}, $a \, M_i(t)$).
 Also, the clusters show a slope along the field due to the second trend (\ref{item:secondtrend}, $b \,\,\, \hat{t} \,\, \hat{x}_i$).
 Finally, PDT exactly identified a cluster of localized trend (\ref{item:thirdtrend}, $\zeta_i(t)$, marked as circles in Fig. \ref{fig:TP}).
 As the results clearly indicate, PDT can identify and group light-curves according to their similarity,
 even though multiple trends are mixed together and the trends are contaminated  by other noise sources such as Poisson noise.

 The identified clusters do not contain any stars  which are intrinsically variable (three transits and two eclipsing binaries).
 This shows that our clustering algorithm is also effectively excluding such unwanted outliers.

\subsection{De-trending Results and Comparison with TFA}
\label{sec:transit}

Here we compare our results to TFA. TFA is one of the particularly successful
de-trending methods \citep{Kovacs2005, Tamuz2005} and it is used by exo-planet searches such as  HATNet
\citep{Bakos2004}. It is therefore a good comparison algorithm for our de-trending algorithm.
 TFA uses a large number of bright stars as a template set while excluding the light-curve being de-trended.
 TFA  does not eliminate potentially dangerous stars, such as the stars which have intrinsic variability,
 from the template set, and it assigns one free parameter per template light-curve.
 In contrast, our algorithm can automatically exclude such intrinsically variable stars and assign one free parameter per cluster of template light-curves.

\begin{figure*}
\begin{center}
        \includegraphics[width=.9\textwidth]{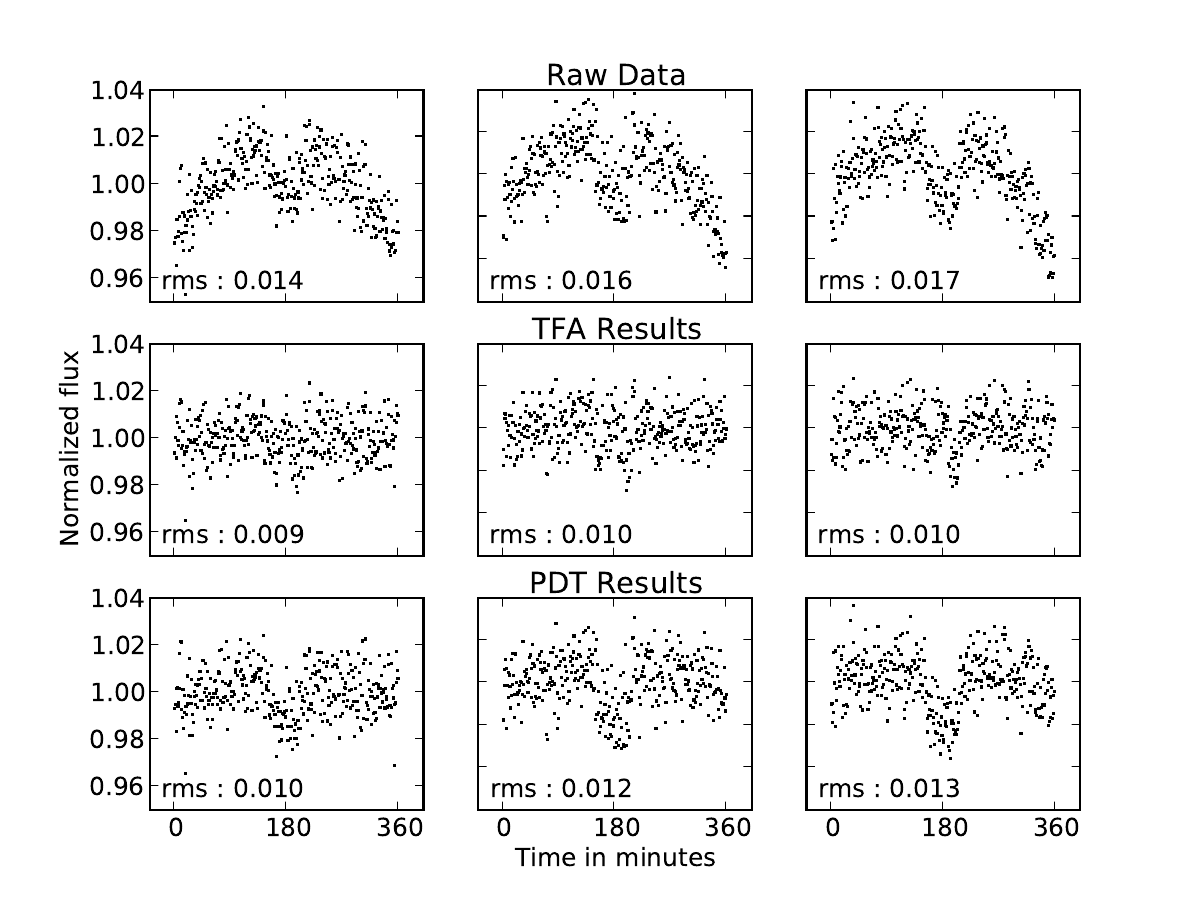}
\end{center}
    \caption
           {
           De-trended results of the simulated three transit events.
           Each column represents each different transit depth of 0.15, 0.20 and 0.25 mag from left to right.
          The top panel is  raw light-curves.
          The middle panel is TFA results, and the bottom panel is PDT results.
          We indicate the rms of each light-curve as well.
           }
    \label{fig:Detrended}
\end{figure*}

 First, we present the de-trended results of three transit signals.
 The top panel of Fig. \ref{fig:Detrended} shows the raw light-curves before any de-trending treatments.
 Each column shows three different transits with different depth (0.015, 0.020 and 0.025 mag from left to right).
 TFA results are shown in the middle panel of Fig. \ref{fig:Detrended},  while PDT results are shown in the bottom panel.
  We used 60 bright stars as a template set for TFA.
 We excluded the  three transits light-curves from the template set because in realistic scenarios,
 it is uncommon for there to be three transit events occurring in the same field and same epoch.
 However, we did not exclude the two eclipsing binaries from the template set for TFA
 because variable stars such as eclipsing binaries are common in the field.
  As the middle panel shows, TFA suppressed each transit signal more than PDT.
 The suppression was mainly caused by the presence of the eclipsing binaries in the template set.
 Because TFA tries to minimize the residual between the target light-curve to be de-trended
 and a linear combination of light-curves from the template set that might contain intrinsic variables
  such as  the two eclipsing binaries in this simulation, it occasionally suppresses
  the intrinsic signals of the target light-curve  by removing any similar signals between the target light-curve and the template set.
 In contrast, results from PDT, which can select template sets that do not  contain the three transits or the two eclipsing binaries,
 show less significant signal depression  and clearer transit signals than TFA results (see bottom panel of Fig. \ref{fig:Detrended}).

Note that one of the eclipsing binaries was phased to the transits to show signal depression effect of TFA.
Such coincidences are not common, but we cannot ignore the probability especially in the case of wide field surveys which  simultaneously  monitor more than several hundred stars. If we exclude the eclipsing binary from the template set, the de-trended results using TFA are almost identical to the results using PDT.

  We performed $\chi^2$ tests comparing de-trended results and  original transit signals for all
  three transits to check how successfully both de-trending algorithms regenerated the intrinsic signals.
 Table \ref{tab:pvalue} shows individual $\chi^2$ values of each transit and $\chi^2$ ratios of TFA to PDT.
 The $\chi^2$ ratio is defined as ${\chi^2_{\rm{TFA}}} / {\chi^2_{\rm{PDT}}}$.
 Therefore, if the $\chi^2$ ratio is bigger than one, it means that
 PDT results are more similar to the original transit signals than TFA results.
 As Table \ref{tab:pvalue} shows,  all three $\chi^2$ ratios are slightly bigger than one.

 \begin{table}
 \begin{center}
 \caption{$\chi^2$ values of each transit and $\chi^2$ ratio of TFA to PDT}
 \begin{tabular}{cccc}
 \hline
 Transit depth &  TFA & PDT & $\chi^2$ ratio \\
 \hline
 0.015 & 1.08 & 1.03 & 1.05 \\
 0.020 & 1.49 & 1.39 & 1.07 \\
 0.025 & 1.69 & 1.50 & 1.13 \\
  \hline
 \label{tab:pvalue}
 \end{tabular}
 \end{center}
 \end{table}

  If the rms contribution from intrinsic signal is significant, such as the two eclipsing binaries in this simulation,
  any method which minimizes the rms to de-trend light-curves will dilute the intrinsic signals.
 This is the critical problem of the rms minimization algorithm and cannot be perfectly overcome as long as we use the minimization approach. One solution to reduce this side effect  is to decrease the number of free parameters (see Sec. \ref{sec:Algorithm})
 and constraint the free parameters to be bigger than or equal to zero (see Sec. \ref{sec:DetermineRemovalTrend}).
 PDT, by construction, has fewer parameters than  TFA because we
 determine one master-trend per one cluster. Also, PDT can constraint the free parameters using quadratic programming.

\begin{figure*}
\begin{center}
        \includegraphics[width=.9\textwidth]{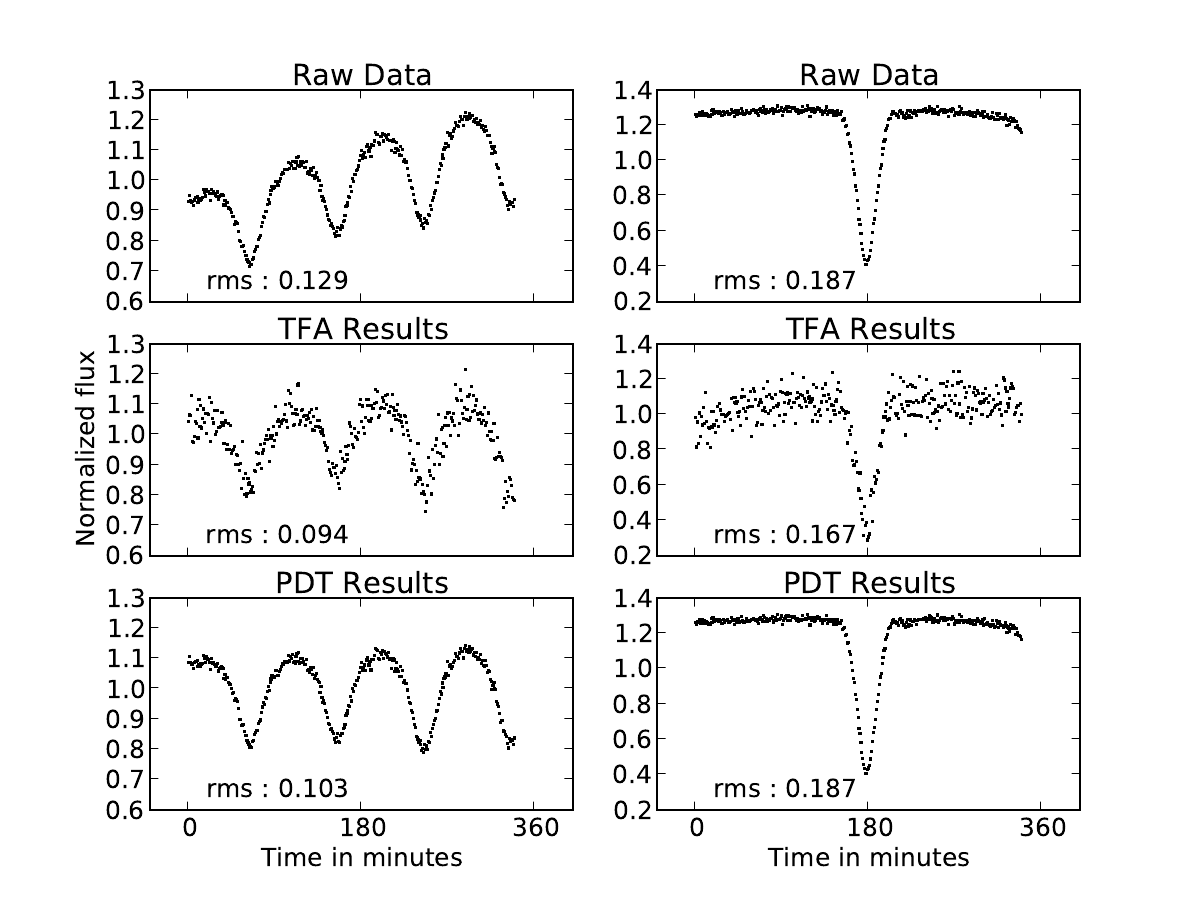}
\end{center}
    \caption
           {De-trended results of the simulated two eclipsing binaries. The top panel is raw light-curves.
           The middle panel is TFA results, and the bottom panel is PDT results.
           We indicate the rms of each light-curve as well.}
    \label{fig:Eclipsing}
\end{figure*}

 Fig. \ref{fig:Eclipsing} shows the de-trended results of the two eclipsing binaries by both TFA and PDT.
The top left panel is the raw light-curve of one eclipsing binary affected by all three trends including localized trend (trend \ref{item:thirdtrend}) and thus the average flux of the light-curve is increasing along time. The top right panel is the raw light-curve of another eclipsing binary affected by only two trends (trend \ref{item:firsttrend} and trend \ref{item:secondtrend}) and thus it
does not show increase of flux because the intrinsic signal is relatively bigger than the two trends. The middle panel is the TFA results, and the bottom panel is the PDT results. In both cases, TFA not only removed the trends but also diluted the intrinsic signals. In contrast, PDT removed only the trends and successfully regenerated the intrinsic signals of two binaries.

 In addition, we indicate the rms of the de-trended light-curves in each panel of Fig. \ref{fig:Detrended} and \ref{fig:Eclipsing}.
 The rms values are always smaller in TFA results than in PDT results because TFA has more adjustable free parameters than PDT has and TFA can set free parameters to be any values including negative values.
 However, as the de-trended results show, smaller rms do not always mean better de-trended results,
 especially when intrinsic signals contribute mainly to rms of light-curves such as the two eclipsing binaries.

Note the TFA has a {\em reconstruction} phase which can greatly improve S/N of periodic signals
with the initial guess of the signal models \citep{Kovacs2005, Kovacs2008}.
Nevertheless, PDT is designed to regenerate any types of intrinsic signals whether they are periodic or not.

\subsection{Second Order Extinction and Other Considerations}
\label{sec:secondANDotherconsideration}

 \subsubsection{Second Order Extinction Test}
 \label{sec:secondorderextinction}

 We now turn our attention to the second order atmospheric extinction
related to colors of stars (\ref{item:fourthtrend} at Sec.\ref{sec:DataDescription}).
  After performing several  simulations with realistic parameters
  (e.g. different field of view from .1 deg$^2$ to 5 deg$^2$, different bands such as B and V,
  different observation durations from one hour to six hours, etc)   that contain both first and second order extinction,
  we found that PDT cannot separate clusters according to  star color.
 The reason is  that  both extinctions depend linearly on airmass,
 and  the first order extinction is much larger than the second order extinction when using  realistic values for the coefficients.
 Therefore, PDT identifies clusters that mainly depend on the first order extinction.

It is worth mentioning that PDT can identify clusters based on colors if we  isolate only the second order extinction.
 We performed another simulation to test this:

 \begin{enumerate}

 \item \label{color:generate} Generate $\sim$500 light-curves that contain only the second order extinction and Poisson noise.
 We extracted B-R colors of stars from USNO B1.0 catalog  within a particular  patch of the sky   ($3 \deg^2$) [$4^h \, 48^m \, 00^s, \, 20^\circ \, 46' \, 20''$], which is the same field of view as in the previous simulation shown in Sec. \ref{sec:DataDescription}.

 \item Apply PDT to the light-curves and identify clusters.

 \end{enumerate}

Table \ref{tab:color} shows the mean and standard deviation values
($\sigma$) of the colors of stars in clusters identified by PDT.
Although some of the clusters (e.g. $C_1$ and $C_2$) could be regarded as clusters of the same trend
because they have similar mean color values,
PDT did a good job of separating bluish cluster ($C_6$) from reddish ($C_1$ to $C_5$) clusters.

 \begin{table}
 \begin{center}
 \caption{Mean and standard deviation values ($\sigma$) of colors of stars in resulting six clusters}
 \begin{tabular}{ccc}
 \hline
 Cluster & mean & $\sigma$ \\
 \hline
$C_1$ & 1.40 & 0.11 \\
$C_2$ & 1.39 & 0.17 \\
$C_3$ & 1.18 & 0.27 \\
$C_4$ & 0.87 & 0.15 \\
$C_5$ & 0.79 & 0.09 \\
$C_6$ & -0.25 & 0.05 \\
 \hline
 \label{tab:color}
 \end{tabular}
 \end{center}
 \end{table}

 \subsubsection{Pure Poisson Noise Case}
 \label{sec:PurePoissonNoise}
  We tested both PDT and TFA with $\sim$500 synthetic light-curves  with pure Poisson noise but no trends.
  We also added three artificial transit events into three individual light-curves. We used 60 bright stars as a template set for TFA.
  Even though  there were no trends, TFA still de-trended the light-curves by using the template stars,
  and it eventually suppressed the intrinsic signals of the transits.
  By contrast, PDT did not  identify any clusters because we exclude  clusters that consist of only Poisson noise (see Sec. \ref{sec:AgglomerativeAlgorithm}).  Consequently, it did not de-trend light-curves and thus did not suppress any intrinsic signals.

Note that Poisson noise is not always the dominant noise source in light-curves.
 The example referred here shows that if there are none-strongly correlated elements (trends) in dataset,
 then PDT will not de-trend the dataset.

\section{Test with Astronomical Datasets}
 \label{sec:TAOSandMMT}

We present now two examples of astronomical datasets. One is from TAOS (The Taiwan-American Occultation Survey) \citep{Lehner2009}, and the other is from an ccoultation survey using Megacam  on the 6.5m Multi-Mirror Telescope (MMT)  \citep{Bianco2009}. Both examples show multiple trends that are well localized on the CCD plane. Such localization of trends could be caused by various noise sources such as airmass, cloud passages, noise of CCD images, telescope vibration, defects of photometry and so on. These localizations often happen with wide field observations.

\subsection{An Example of TAOS Dataset}
 \label{sec:TAOSexam}

The scientific goal of TAOS \citep{Zhang2008ApJL} is to detect km-sized Kuiper Belt Objects \citep{Luu2002} at a distance of Neptune or beyond. TAOS data usually suffer from low S/N and systematic trends due to the small telescope size (four 50cm telescopes), noise of CCD images, defects of photometry and unstable local weather (e.g. cloud passages). The field of view of the TAOS telescopes is 3 deg${}^2$ and the sampling rate is 5 Hz. We chose one sample set of light-curves generated by the TAOS photometry pipeline \citep{Zhang2009} and de-trended the light-curves using PDT. The total observation time of the light-curves was 1.5 hours.

\begin{figure*}
\begin{center}
\begin{minipage}[c]{8cm}
        \includegraphics[width=1.0\textwidth]{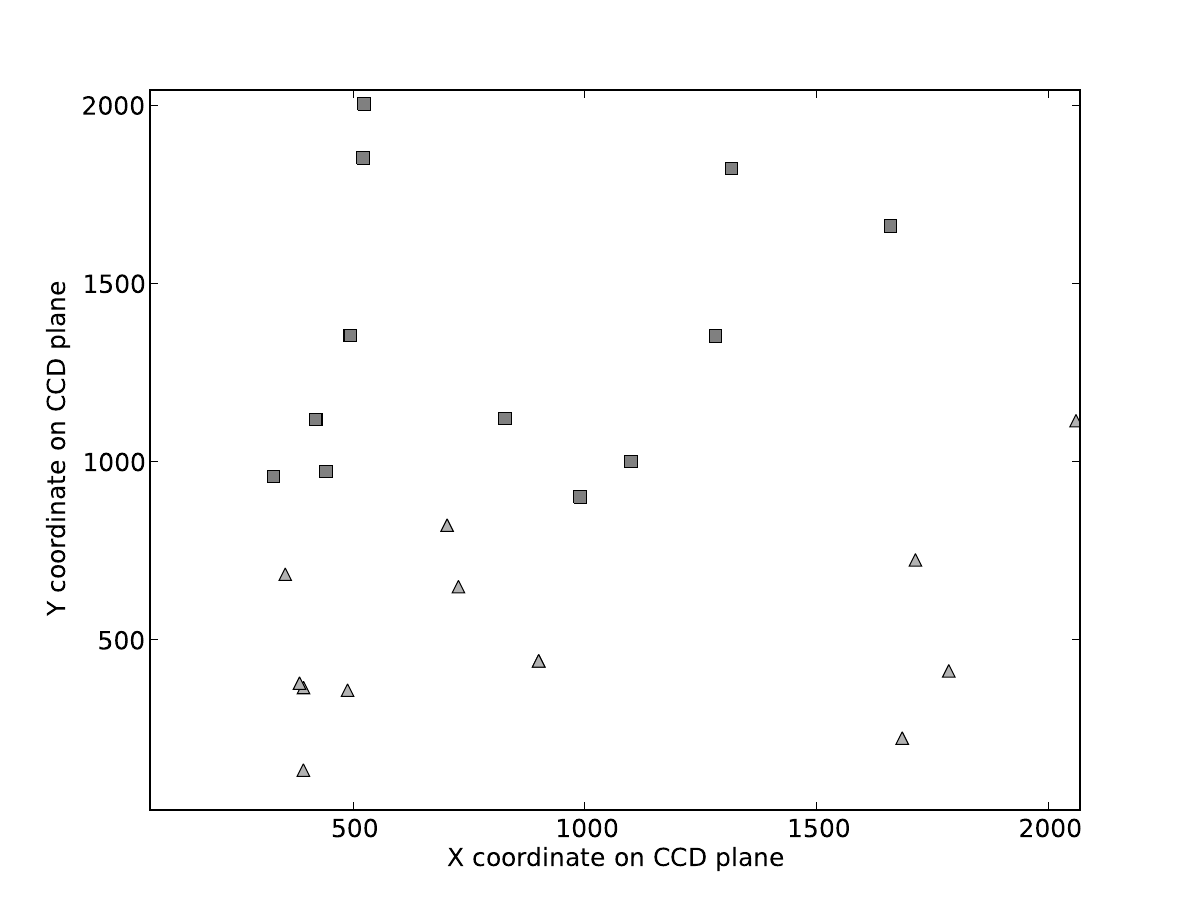}
\end{minipage}
\begin{minipage}[c]{8cm}
        \includegraphics[width=1.0\textwidth]{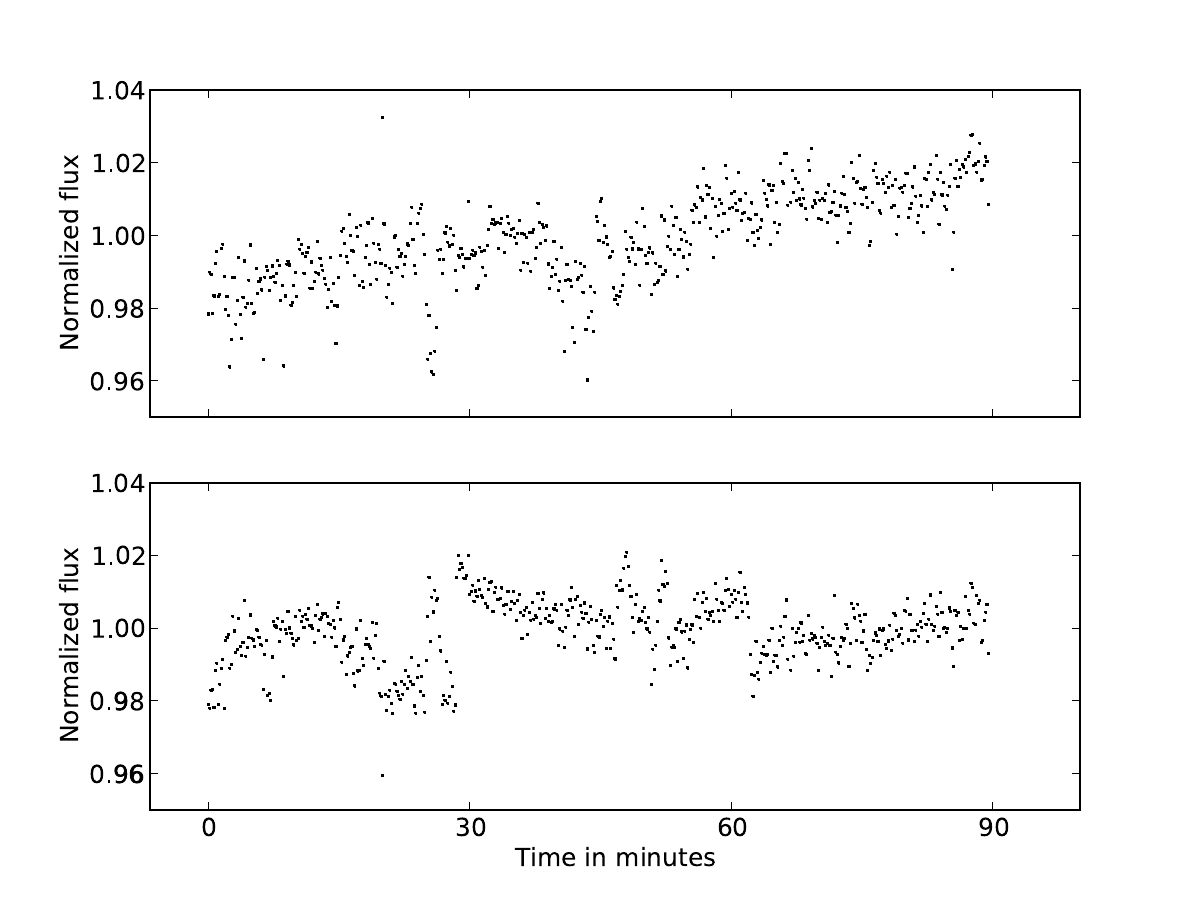}
\end{minipage}
\\
\begin{minipage}[c]{8cm}
        \includegraphics[width=1.0\textwidth]{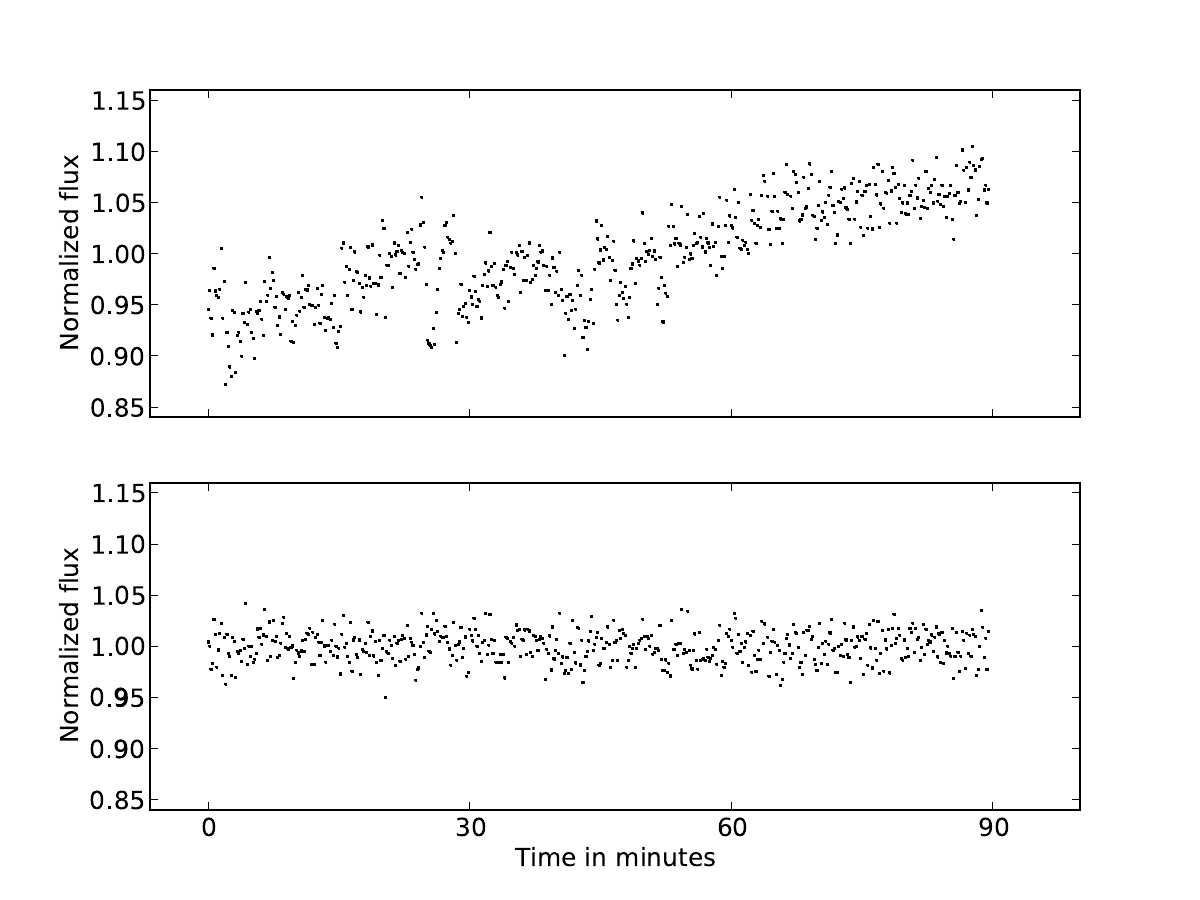}
\end{minipage}
\begin{minipage}[c]{8cm}
        \includegraphics[width=1.0\textwidth]{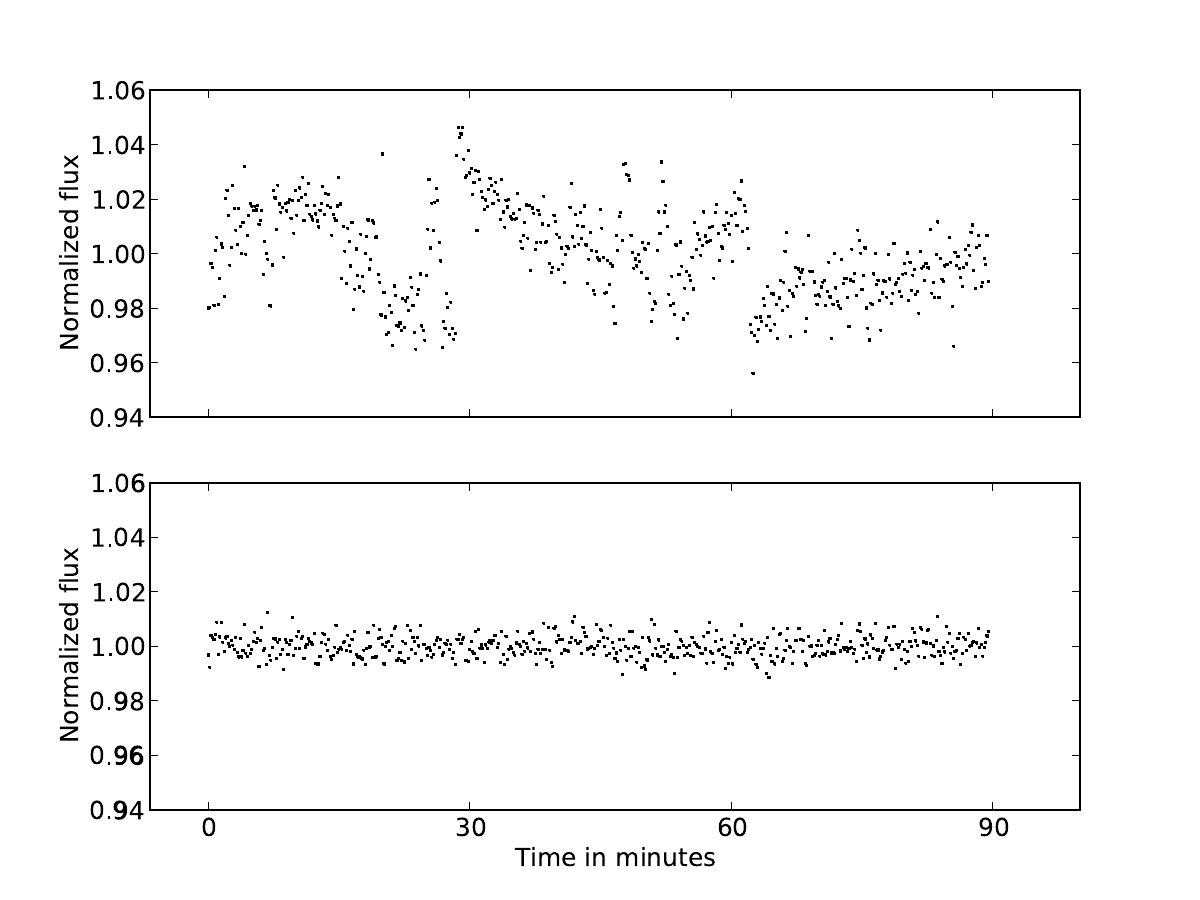}
\end{minipage}
\end{center}
    \caption
           {An example of TAOS dataset. Top left : Position of stars in identified two clusters.  x(y)-axis is the x(y)-coordinate of stars on the CCD plane.  Different shapes indicate different clusters. Top right : Determined two master-trends. Bottom left and bottom right : Two example light-curves before and after de-trending. Upper panels are before de-trending and lower panels are after de-trending.}
    \label{fig:TAOS}
\end{figure*}

Fig. \ref{fig:TAOS} shows the determined master-trends and examples of de-trended light-curves. The top left panel shows the position of stars in identified clusters on the CCD plane. Different shapes indicate different clusters. The clusters are localized on the CCD plane due to unstable local weather, noise of CCD images and defects of photometry. The bottom two panels show example light-curves of two non-variable stars. The upper light-curves of the two bottom panels are before de-trending and the lower light-curves are after de-trending. As the results show, PDT removed trends from both light-curves.

\subsection{An Example of Megacam Dataset}
 \label{sec:MMTSexam}

We also applied PDT to a dataset obtained using Megacam \citep{McLeod1998}
at the MMT at Mount Hopkins, Arizona. Megacam is a mosaic CCD which consists of 36 chips.
The size of each CCD is 2K by 4K and the field of view is 24$'$ x 24$'$.
Megacam was used in {\em continuous-readout} mode achieving 200Hz sampling rate
in order to detect stellar occultations caused by Kuiper Belt Objects \citep{Bianco2009}.
Due to the high sampling rate, telescope vibrations, defects of photometry and  the readout technique,
these Megacam data show strong trends.
The total observation time of the selected dataset was 15 minutes.

\begin{figure*}
\begin{center}
\begin{minipage}[c]{8cm}
        \includegraphics[width=1.0\textwidth]{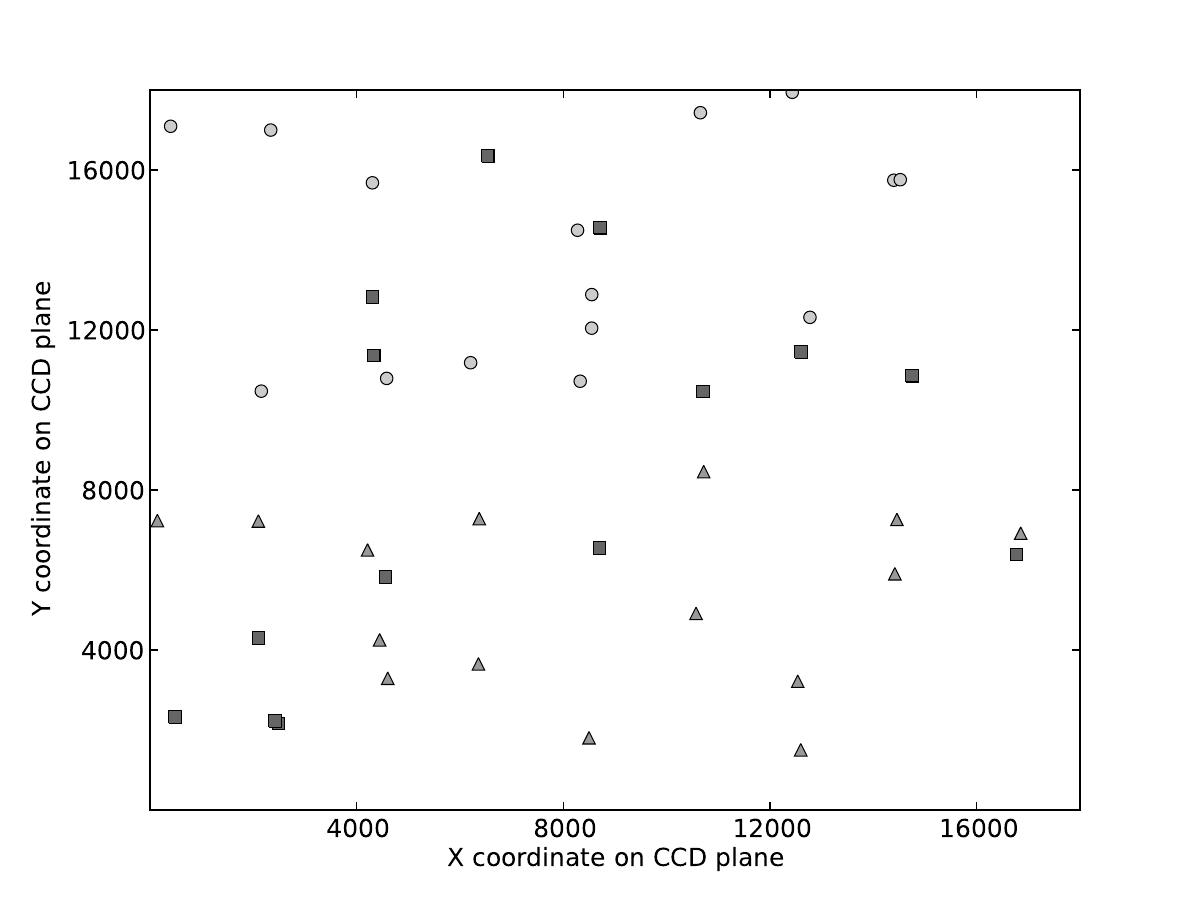}
\end{minipage}
\begin{minipage}[c]{8cm}
        \includegraphics[width=1.0\textwidth]{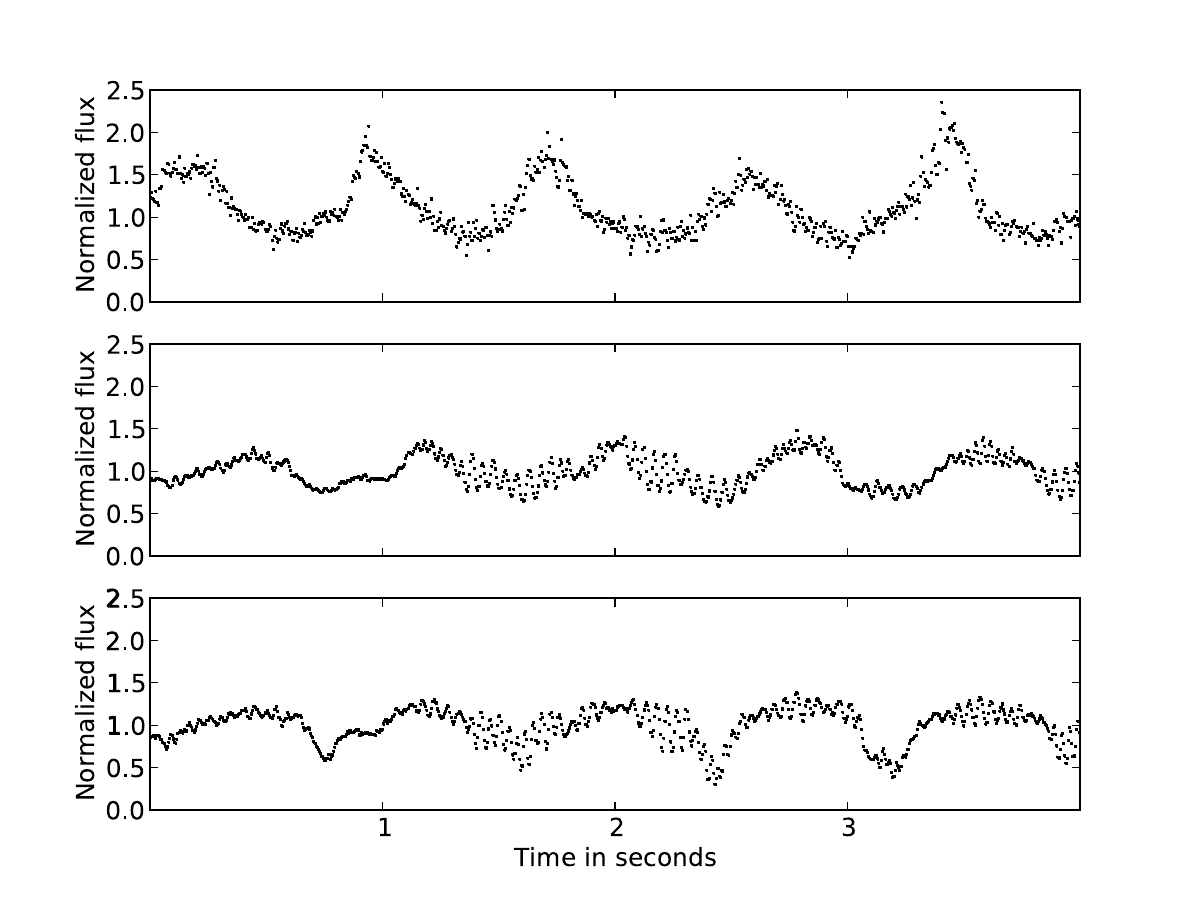}
\end{minipage}
\\
\begin{minipage}[c]{8cm}
        \includegraphics[width=1.0\textwidth]{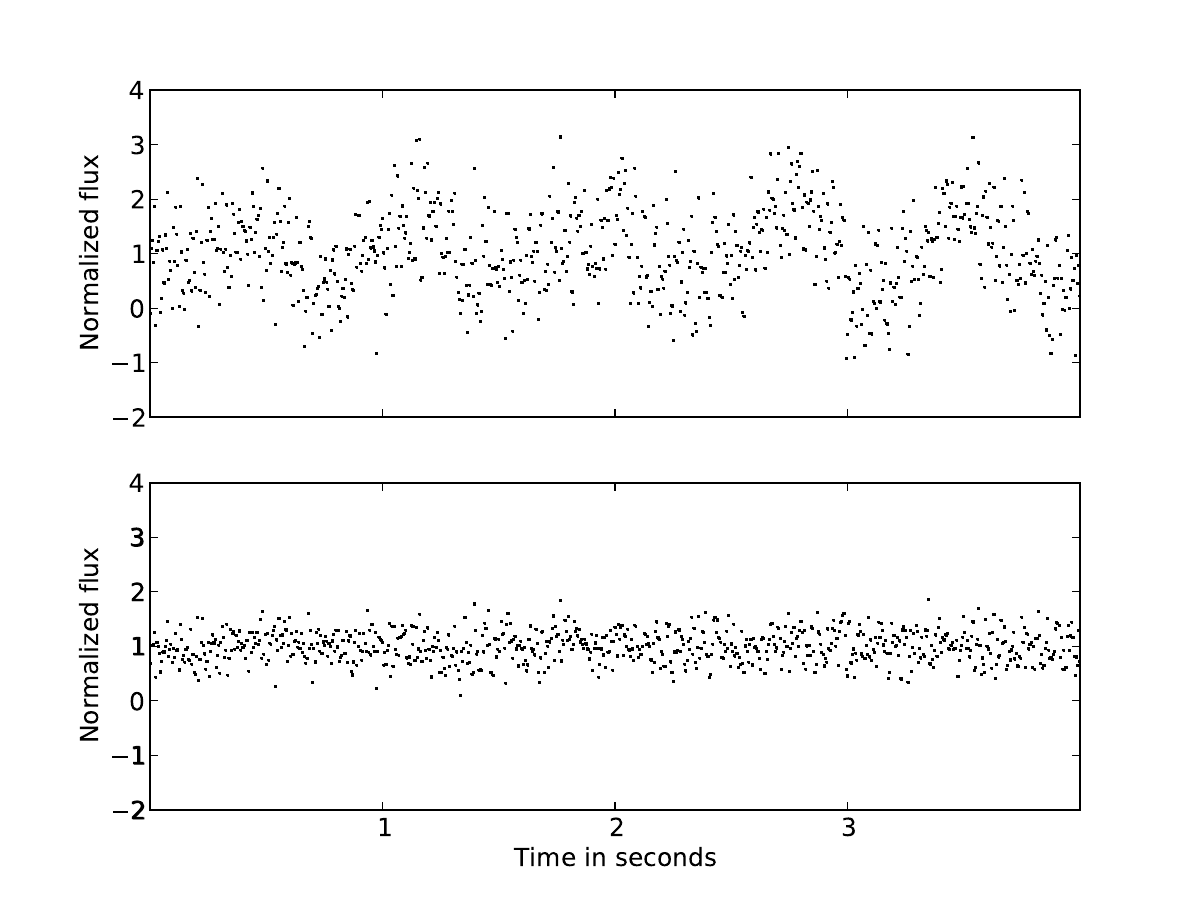}
\end{minipage}
\begin{minipage}[c]{8cm}
        \includegraphics[width=1.0\textwidth]{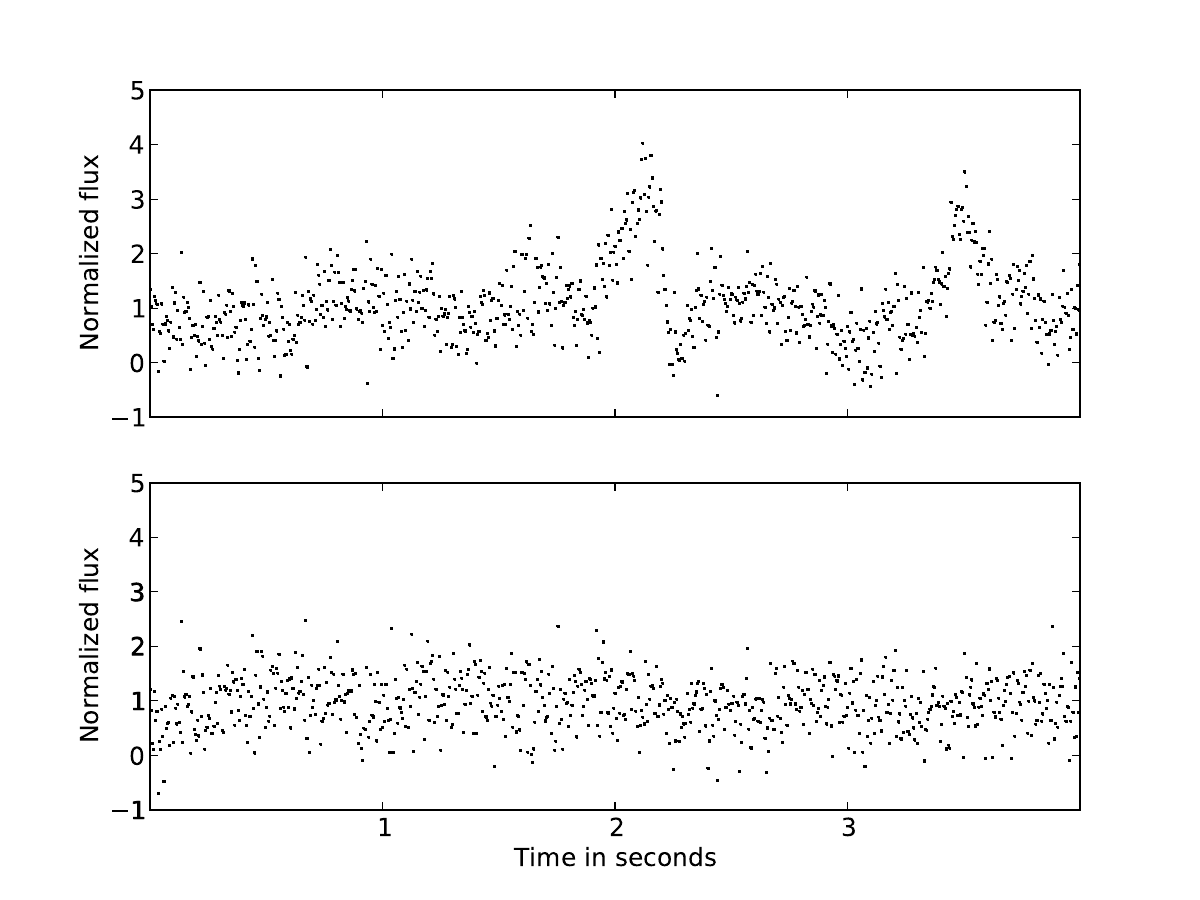}
\end{minipage}
\end{center}
    \caption
           {An example of Megacam dataset. Top left : Position of stars in identified three clusters.  x(y)-axis is the x(y)-coordinate of stars on the CCD plane.  Different shapes indicate different clusters. Top right : Determined three master-trends. Bottom left and bottom right : Two example light-curves before and after de-trending. Upper panels are before de-trending and lower panels are after de-trending.}
    \label{fig:MMT}
\end{figure*}

The top left panel of Fig. \ref{fig:MMT} shows the position of stars in identified clusters.
Different shapes indicate different clusters.
The top right panel shows the determined master-trends.
We magnified a part of the light-curves ($\sim$5 seconds) to clearly show the trends.
The bottom two panels show two example light-curves of non-variable stars before and after de-trending.

As the figure shows, two clusters marked as circles and triangles are localized on the CCD plane.
In our analysis, we found that often the clusters were divided along the
horizontal half divide (e.g. clusters marked as circles and triangles in Fig. \ref{fig:MMT}),
and that can be attributed to details of the readout mode,
but we also found cases where the clustering
that crossed over the horizontal divide (e.g. a cluster marked as squares in \ref{fig:MMT}).
The trends are likely due to a combination of weather patterns,
photometry and the way the CCD was read out \citep{Bianco2009}.

\section{Notes and Future Work}
\label{sec:NoteFutureWork}

  A weakness of PDT is that it cannot remove  trends  that are manifested in just a few light-curves and are not highly correlated.  For example, moving asteroids or satellites could result in an increase and decrease of the estimated flux of a few background stars in the neighborhood of the track.  These trends are out of phase throughout  the light-curves  because  the asteroids or satellites are moving across the field. For these reasons, strongly trended light-curves are not  highly correlated and thus PDT cannot group them into clusters. We are planning to handle this phase-shift of trends in a future version of PDT.

We are also applying PDT to astronomical datasets, e.g. TAOS and MMT,
in order to detect various transient events such as KBO occultations, flare stars,
micro-lensing events and exo-planet transits.

\section{Conclusion}
\label{sec:Conclusion}

  In this paper, we presented the Photometric De-Trending algorithm (PDT), a new de-trending algorithm.
 We first determined the trends by constructing a hierarchical tree based on the similarity matrix.
 Elements of the similarity matrix are the Pearson correlation values of all pairs of light-curves.
 After that, a bottom-up merging algorithm was applied to the constructed tree in order to identify subsets of light-curves that we call clusters.
 At each step of the merging process, we tested the normality of the subsets and determined where to stop.
 By means of the normality test, we could select reliable clusters of trends.
 For each cluster, we determined one representative master-trend by weighted sum  of the normalized light-curves.
 This procedure greatly constrained the number of free parameters to be calculated,
 and thus  showed less significant signal depression than other de-trending algorithms such as Trend Filtering Algorithm (TFA).
 Finally, in order to remove the trends from individual light-curves,
 we used quadratic programming to minimize the residual  between each target light-curve and the determined master-trends.
 Note that PDT is designed to remove only the fluctuations that are common among stars.
 If the fluctuations are unique to an individual star, the fluctuations will be preserved.

 We performed several simulations of synthetic light-curves with different initial parameters such as  total duration of observation,
 transit duration, field of view, exposure time etc,  to test PDT and showed some of the simulation results in this paper.

 First, we tested PDT with $\sim$500 synthetic light-curves  that contain the first order atmospheric extinction (airmass), artificial trends,
 Poisson noise and events (three transits and two eclipsing binaries).
 We applied PDT to these synthetic light-curves in order to determine trends  and to regenerate the inserted events.
 PDT successfully identified multiple clusters of different  trends which were the mixture of different trends and noise.
 These identified clusters well represented the overall characteristic of the trends through the field.
 We compared de-trended results of PDT with one another de-trending algorithm (TFA).
 PDT results were an improvement over TFA results, especially
 when 1) the dataset contains intrinsic variables  that would be included in template set of TFA
 or 2) the rms contribution from the intrinsic signals is significant.

We also tested PDT with $\sim$500 synthetic light-curves
that contain color dependent second order extinction and Poisson noise.
Trends appearing in light-curves can be slightly different due to differences in color.
We found that PDT can identify clusters according to color.
However, in realistic scenarios, it is not easy to isolate only the second order extinction  because,
even if one correctly removes the first order extinction for all stars,
there exist other various noise sources which dilute trends caused by the second order extinction.

 In the case of dataset of random fluctuations (e.g. pure Poisson noise),
 which does not have any trends, we do not need to de-trend the dataset.
 PDT can distinguish the light-curves of random fluctuations
 using the characteristics of the distribution of correlation coefficients.
 Therefore, PDT does not de-trend these light-curves and thus preserves any intrinsic signals.

 Examples of two astronomical datasets are also presented.
They show multiple trends in the field caused by various noise sources such as
airmass, cloud passages, telescope vibration, defects of photometry and so on.
PDT performed well and removed trends that appeared in the datasets.

  In this paper, we show the simulation results of wide field data only.
 However, PDT can be applied to narrow field data as well if there are enough stars in the field ($\sim$ a few hundreds).
 In addition, PDT is useful to extract global trends that can represent the overall  characteristics of a dataset.
 The extracted trends can give a general idea  of how much the data are contaminated by the trends.

 The software package of PDT is be provided at
 \href{http://timemachine.iic.harvard.edu/detrending01/}{http://timemachine.iic.harvard.edu}.

\section*{Acknowledgements}
  This work is supported by the Korea Research Foundation.
  Y.-I. Byun also acknowledges the grant of KRF-2007-C00020.
  We thank R. Reid, R. Dave, G. Wachman and D. Preston  at \href{http://iic.harvard.edu}{the Harvard, Initiative in Innovative Computing} (IIC),
  and A. W. Blocker   at  \href{http://www.stat.harvard.edu}{the Harvard University Department of Statistics}
   for comments and suggestions on this paper.   We also thank
   \href{http://cfa-www.harvard.edu}{the Harvard-Smithsonian Center for Astrophysics} and IIC
   for providing computing facilities and research space.
The simulations and the de-trending of datasets in this paper were run on the
\href{http://hptc.fas.harvard.edu/}{Odyssey cluster} supported by the FAS Research Computing Group at the Harvard.

\bibliography{PDT}{}

\begin{thebibliography}{}

\bibitem[\protect\citeauthoryear{{Akerlof}, {Amrose}, {Balsano}, {Bloch},
  {Casperson}, {Fletcher}, {Gisler}, {Hills}, {Kehoe}, {Lee}, {Marshall},
  {McKay}, {Pawl}, {Schaefer}, {Szymanski} \& {Wren}}{{Akerlof}
  et~al.}{2000}]{Akerlof2000}
{Akerlof} C.,  {Amrose} S.,  {Balsano} R.,  {Bloch} J.,  {Casperson} D.,
  {Fletcher} S.,  {Gisler} G.,  {Hills} J.,  {Kehoe} R.,  {Lee} B.,  {Marshall}
  S.,  {McKay} T.,  {Pawl} A.,  {Schaefer} J.,  {Szymanski} J.,    {Wren} J.,
  2000, AJ, 119, 1901

\bibitem[\protect\citeauthoryear{{Alonso}, {Brown}, {Torres}, {Latham},
  {Sozzetti}, {Mandushev}, {Belmonte}, {Charbonneau}, {Deeg}, {Dunham},
  {O'Donovan} \& {Stefanik}}{{Alonso} et~al.}{2004}]{Alonso2004}
{Alonso} R.,  {Brown} T.~M.,  {Torres} G.,  {Latham} D.~W.,  {Sozzetti} A.,
  {Mandushev} G.,  {Belmonte} J.~A.,  {Charbonneau} D.,  {Deeg} H.~J.,
  {Dunham} E.~W.,  {O'Donovan} F.~T.,    {Stefanik} R.~P.,  2004, ApJL, 613,
  L153

\bibitem[\protect\citeauthoryear{{Anderson}}{{Anderson}}{1996}]{Anderson1996}
{Anderson} T.~W.,  1996, Statistical Science, 11, 20

\bibitem[\protect\citeauthoryear{{Anderson} \& {Darling}}{{Anderson} \&
  {Darling}}{1952}]{Anderson1952}
{Anderson} T.~W.,  {Darling} D.~A.,  1952, Annals of Mathematical Statistics,
  23, 193

\bibitem[\protect\citeauthoryear{{Bakos}, {Noyes}, {Kov{\'a}cs}, {Stanek},
  {Sasselov} \& {Domsa}}{{Bakos} et~al.}{2004}]{Bakos2004}
{Bakos} G.,  {Noyes} R.~W.,  {Kov{\'a}cs} G.,  {Stanek} K.~Z.,  {Sasselov}
  D.~D.,    {Domsa} I.,  2004, PASP, 116, 266

\bibitem[\protect\citeauthoryear{{Bakos}, {Noyes}, {Kov{\'a}cs}, {Latham},
  {Sasselov}, {Torres}, {Fischer}, {Stefanik}, {Sato}, {Johnson}, {P{\'a}l},
  {Marcy}, {Butler}, {Esquerdo} \& et al.}{{Bakos} et~al.}{2007}]{Bakos2007}
{Bakos} G.~{\'A}.,  {Noyes} R.~W.,  {Kov{\'a}cs} G.,  {Latham} D.~W.,
  {Sasselov} D.~D.,  {Torres} G.,  {Fischer} D.~A.,  {Stefanik} R.~P.,  {Sato}
  B.,  {Johnson} J.~A.,  {P{\'a}l} A.,  {Marcy} G.~W.,  {Butler} R.~P.,
  {Esquerdo} G.~A.,    et al. 2007, ApJ, 656, 552

\bibitem[\protect\citeauthoryear{{Bianco}, {Protopapas}, {McLeod}, {Alcock},
  {Holman} \& {Lehner}}{{Bianco} et~al.}{2009}]{Bianco2009}
{Bianco} F.~B.,  {Protopapas} P.,  {McLeod} B.~A.,  {Alcock} C.~R.,  {Holman}
  M.~J.,    {Lehner} M.~J.,  2009, ArXiv e-prints

\bibitem[\protect\citeauthoryear{{Bowley}}{{Bowley}}{1928}]{Bowley1928}
{Bowley} A.~L.,  1928, Journal of the American Statistical Association, 23, 31

\bibitem[\protect\citeauthoryear{{Burke}, {McCullough}, {Valenti}, {Long},
  {Johns-Krull}, {Machalek}, {Janes}, {Taylor}, {Fleenor}, {Foote}, {Gary},
  {Garc{\'{\i}}a-Melendo}, {Gregorio} \& {Vanmunster}}{{Burke}
  et~al.}{2008}]{Burke2008}
{Burke} C.~J.,  {McCullough} P.~R.,  {Valenti} J.~A.,  {Long} D.,
  {Johns-Krull} C.~M.,  {Machalek} P.,  {Janes} K.~A.,  {Taylor} B.,  {Fleenor}
  M.~L.,  {Foote} C.~N.,  {Gary} B.~L.,  {Garc{\'{\i}}a-Melendo} E.,
  {Gregorio} J.,    {Vanmunster} T.,  2008, ApJ, 686, 1331

\bibitem[\protect\citeauthoryear{D'Agostino \& Stephens}{D'Agostino \&
  Stephens}{1986}]{D'Agostino1993}
D'Agostino R.~B.,  Stephens M.~A.,  1986, {Goodness-of-fit techniques}.
Marcel Dekker, Inc., New York, NY, USA, p. 107

\bibitem[\protect\citeauthoryear{{Daniels} \& {Giraud-Carrier}}{{Daniels} \&
  {Giraud-Carrier}}{2006}]{Daniels2006}
{Daniels} K.,  {Giraud-Carrier} C.,  2006, in ICMLA '06: Proceedings of the 5th
  International Conference on Machine Learning and Applications {Learning the
  Threshold in Hierarchical Agglomerative Clustering}.
IEEE Computer Society, Washington, DC, USA, pp 270--278

\bibitem[\protect\citeauthoryear{{de Hoon}, {Imoto}, {Nolan} \& {Miyano}}{{de
  Hoon} et~al.}{2004}]{deHoon2004}
{de Hoon} M.~J.~L.,  {Imoto} S.,  {Nolan} J.,    {Miyano} S.,  2004,
  BIOINFORMATICS, 20, 1453

\bibitem[\protect\citeauthoryear{{Ester}, {Kriegel}, {Sander} \& {Xu}}{{Ester}
  et~al.}{1996}]{Ester1996}
{Ester} M.,  {Kriegel} H.~P.,  {Sander} J.,    {Xu} X.,  1996, in Proceedings
  of 2nd International Conference on Knowledge Discovery and Data Mining
  (KDD-96) {A Density-Based Algorithm for Discovering Clusters in Large Spatial
  Databases with Noise}.
Portland, Oregon, USA, pp 226--231

\bibitem[\protect\citeauthoryear{{Everett} \& {Howell}}{{Everett} \&
  {Howell}}{2001}]{Everett2001}
{Everett} M.~E.,  {Howell} S.~B.,  2001, PASP, 113, 1428

\bibitem[\protect\citeauthoryear{{Everett}, {Howell}, {van Belle} \&
  {Ciardi}}{{Everett} et~al.}{2002}]{Everett2002}
{Everett} M.~E.,  {Howell} S.~B.,  {van Belle} G.~T.,    {Ciardi} D.~R.,  2002,
  PASP, 114, 656

\bibitem[\protect\citeauthoryear{{Fisher}}{{Fisher}}{1915}]{Fisher1915}
{Fisher} R.~A.,  1915, Biometrika, 10, 507

\bibitem[\protect\citeauthoryear{{Ghosh}}{{Ghosh}}{1966}]{Ghosh1966}
{Ghosh} B.~K.,  1966, Biometrika, 53, 258

\bibitem[\protect\citeauthoryear{{Gilliland} \& {Brown}}{{Gilliland} \&
  {Brown}}{1988}]{Gilliland1988}
{Gilliland} R.~L.,  {Brown} T.~M.,  1988, PASP, 100, 754

\bibitem[\protect\citeauthoryear{{Goldfarb} \& {Idnani}}{{Goldfarb} \&
  {Idnani}}{1983}]{Goldfarb1983}
{Goldfarb} D.,  {Idnani} A.,  1983, Mathematical Programming, 27, 1

\bibitem[\protect\citeauthoryear{{Hartigan} \& {Wong}}{{Hartigan} \&
  {Wong}}{1979}]{Hartigan1979}
{Hartigan} J.~A.,  {Wong} M.~A.,  1979, Applied Statistics, 28, 100

\bibitem[\protect\citeauthoryear{{Hotelling}}{{Hotelling}}{1953}]{Hotelling195%
3}
{Hotelling} H.,  1953, Journal of the Royal Statistical Society. Series B
  (Statistical Methodology), 15, 193

\bibitem[\protect\citeauthoryear{{Howell} \& {Jacoby}}{{Howell} \&
  {Jacoby}}{1986}]{Howell1986}
{Howell} S.~B.,  {Jacoby} G.~H.,  1986, PASP, 98, 802

\bibitem[\protect\citeauthoryear{{Jain}, {Murty} \& {Flynn}}{{Jain}
  et~al.}{1999}]{Jain1999}
{Jain} A.~K.,  {Murty} M.~N.,    {Flynn} P.~J.,  1999, ACM Computing Surveys,
  31, 264

\bibitem[\protect\citeauthoryear{{Kim} \& {Shevlyakov}}{{Kim} \&
  {Shevlyakov}}{2008}]{Kim2008}
{Kim} K.,  {Shevlyakov} G.,  2008, Signal Processing Magazine, IEEE, Vol. 25,
  No. 2, pp 102--113

\bibitem[\protect\citeauthoryear{{Kjeldsen} \& {Frandsen}}{{Kjeldsen} \&
  {Frandsen}}{1992}]{Kjeldsen1992}
{Kjeldsen} H.,  {Frandsen} S.,  1992, PASP, 104, 413

\bibitem[\protect\citeauthoryear{{Kov{\'a}cs}, {Bakos} \& {Noyes}}{{Kov{\'a}cs}
  et~al.}{2005}]{Kovacs2005}
{Kov{\'a}cs} G.,  {Bakos} G.,    {Noyes} R.~W.,  2005, MNRAS, 356, 557

\bibitem[\protect\citeauthoryear{{Kovacs} \& {Bakos}}{{Kovacs} \&
  {Bakos}}{2008}]{Kovacs2008}
{Kovacs} G.,  {Bakos} G.~A.,  2008, Communications in Asteroseismology, 157, 82

\bibitem[\protect\citeauthoryear{{Landolt}}{{Landolt}}{1992}]{Landolt1992}
{Landolt} A.~U.,  1992, AJ, 104, 340

\bibitem[\protect\citeauthoryear{{Lehner}, {Wen}, {Wang}, {Marshall},
  {Schwamb}, {Zhang}, {Bianco}, {Giammarco}, {Porrata}, {Alcock}, {Axelrod},
  {Byun}, {Chen}, {Cook}, {Dave}, {King}, {Lee}, {Lin}, {Wang}, {Rice} \& {de
  Pater}}{{Lehner} et~al.}{2009}]{Lehner2009}
{Lehner} M.~J.,  {Wen} C.-Y.,  {Wang} J.-H.,  {Marshall} S.~L.,  {Schwamb}
  M.~E.,  {Zhang} Z.-W.,  {Bianco} F.~B.,  {Giammarco} J.,  {Porrata} R.,
  {Alcock} C.,  {Axelrod} T.,  {Byun} Y.-I.,  {Chen} W.~P.,  {Cook} K.~H.,
  {Dave} R.,  {King} S.-K.,  {Lee} T.,  {Lin} H.-C.,  {Wang} S.-Y.,  {Rice}
  J.~A.,    {de Pater} I.,  2009, PASP, 121, 138

\bibitem[\protect\citeauthoryear{{Luu} \& {Jewitt}}{{Luu} \&
  {Jewitt}}{2002}]{Luu2002}
{Luu} J.~X.,  {Jewitt} D.~C.,  2002, ARA\&A, 40, 63

\bibitem[\protect\citeauthoryear{{Mandel} \& {Agol}}{{Mandel} \&
  {Agol}}{2002}]{Mandel2002}
{Mandel} K.,  {Agol} E.,  2002, ApJL, 580, L171

\bibitem[\protect\citeauthoryear{{McCullough}, {Stys}, {Valenti}, {Fleming},
  {Janes} \& {Heasley}}{{McCullough} et~al.}{2005}]{McCullough2005}
{McCullough} P.~R.,  {Stys} J.~E.,  {Valenti} J.~A.,  {Fleming} S.~W.,  {Janes}
  K.~A.,    {Heasley} J.~N.,  2005, PASP, 117, 783

\bibitem[\protect\citeauthoryear{{McLeod}, {Gauron}, {Geary}, {Ordway} \&
  {Roll}}{{McLeod} et~al.}{1998}]{McLeod1998}
{McLeod} B.~A.,  {Gauron} T.~M.,  {Geary} J.~C.,  {Ordway} M.~P.,    {Roll}
  J.~B.,  1998, in {D'Odorico} S.,  ed., Society of Photo-Optical
  Instrumentation Engineers (SPIE) Conference Series Vol.~3355 of Society of
  Photo-Optical Instrumentation Engineers (SPIE) Conference Series, {Megacam:
  paving the focal plane of the MMT with silicon}.
pp 477--486

\bibitem[\protect\citeauthoryear{{Monet}, {Levine}, {Canzian}, {Ables}, {Bird},
  {Dahn}, {Guetter}, {Harris}, {Henden}, {Leggett}, {Levison}, {Luginbuhl},
  {Martini}, {Monet}, {Munn}, {Pier}, {Rhodes}, {Riepe} \& et al.}{{Monet}
  et~al.}{2003}]{Monet2003}
{Monet} D.~G.,  {Levine} S.~E.,  {Canzian} B.,  {Ables} H.~D.,  {Bird} A.~R.,
  {Dahn} C.~C.,  {Guetter} H.~H.,  {Harris} H.~C.,  {Henden} A.~A.,  {Leggett}
  S.~K.,  {Levison} H.~F.,  {Luginbuhl} C.~B.,  {Martini} J.,  {Monet}
  A.~K.~B.,  {Munn} J.~A.,  {Pier} J.~R.,  {Rhodes} A.~R.,  {Riepe} B.,    et
  al. 2003, AJ, 125, 984

\bibitem[\protect\citeauthoryear{{Ng} \& {Han}}{{Ng} \& {Han}}{1994}]{Ng1994}
{Ng} R.~T.,  {Han} J.,  1994, in Bocca J.,  Jarke M.,   Zaniolo C.,  eds, 20th
  International Conference on Very Large Data Bases, September 12--15, 1994,
  Santiago, Chile proceedings {Efficient and Effective Clustering Methods for
  Spatial Data Mining}.
Morgan Kaufmann Publishers, Los Altos, CA 94022, USA, pp 144--155

\bibitem[\protect\citeauthoryear{{Paczynski} \& {Pojmanski}}{{Paczynski} \&
  {Pojmanski}}{2000}]{Paczynski2000}
{Paczynski} B.,  {Pojmanski} G.,  2000, in Bulletin of the American
  Astronomical Society Vol.~32 of Bulletin of the American Astronomical
  Society, {Monitoring All Sky for Variability}.
p.~687

\bibitem[\protect\citeauthoryear{{P{\'a}l}, {Bakos}, {Torres}, {Noyes},
  {Latham}, {Kov{\'a}cs}, {Marcy}, {Fischer}, {Butler}, {Sasselov}, {Sip{\H
  o}cz}, {Esquerdo}, {Kov{\'a}cs}, {Stefanik}, {L{\'a}z{\'a}r}, {Papp} \&
  {S{\'a}ri}}{{P{\'a}l} et~al.}{2008}]{Pal2008}
{P{\'a}l} A.,  {Bakos} G.~{\'A}.,  {Torres} G.,  {Noyes} R.~W.,  {Latham}
  D.~W.,  {Kov{\'a}cs} G.,  {Marcy} G.~W.,  {Fischer} D.~A.,  {Butler} R.~P.,
  {Sasselov} D.~D.,  {Sip{\H o}cz} B.,  {Esquerdo} G.~A.,  {Kov{\'a}cs} G.,
  {Stefanik} R.,  {L{\'a}z{\'a}r} J.,  {Papp} I.,    {S{\'a}ri} P.,  2008, ApJ,
  680, 1450

\bibitem[\protect\citeauthoryear{{Pigulski} \& {Pojma{\'n}ski}}{{Pigulski} \&
  {Pojma{\'n}ski}}{2008}]{Pigulski2008}
{Pigulski} A.,  {Pojma{\'n}ski} G.,  2008, A\&A, 477, 907

\bibitem[\protect\citeauthoryear{{Pojmanski}}{{Pojmanski}}{2005}]{Pojmanski200%
5}
{Pojmanski} G.,  2005, VizieR On-line Data Catalog: J/other/AcA/50.177.~
  Originally published in: 2000AcA....50..177P, 50, 5001

\bibitem[\protect\citeauthoryear{{Pollacco}, {Skillen}, {Collier Cameron},
  {Loeillet}, {Stempels}, {Bouchy}, {Gibson}, {Hebb}, {H{\'e}brard}, {Joshi},
  {McDonald}, {Smalley}, {Smith}, {Street} \& et al.}{{Pollacco}
  et~al.}{2008}]{Pollacco2008}
{Pollacco} D.,  {Skillen} I.,  {Collier Cameron} A.,  {Loeillet} B.,
  {Stempels} H.~C.,  {Bouchy} F.,  {Gibson} N.~P.,  {Hebb} L.,  {H{\'e}brard}
  G.,  {Joshi} Y.~C.,  {McDonald} I.,  {Smalley} B.,  {Smith} A.~M.~S.,
  {Street} R.~A.,    et al. 2008, MNRAS, 385, 1576

\bibitem[\protect\citeauthoryear{{Schmidt}}{{Schmidt}}{1991}]{Schmidt1991}
{Schmidt} E.~G.,  1991, AJ, 102, 1766

\bibitem[\protect\citeauthoryear{{Schmidt}, {Langan}, {Rogalla} \&
  {Thacker-Lynn}}{{Schmidt} et~al.}{2007}]{Schmidt2007}
{Schmidt} E.~G.,  {Langan} S.,  {Rogalla} D.,    {Thacker-Lynn} L.,  2007, AJ,
  133, 665

\bibitem[\protect\citeauthoryear{{Stalin}, {Hegde}, {Sahu}, {Parihar},
  {Anupama}, {Bhatt} \& {Prabhu}}{{Stalin} et~al.}{2008}]{Stalin2008}
{Stalin} C.~S.,  {Hegde} M.,  {Sahu} D.~K.,  {Parihar} P.~S.,  {Anupama} G.~C.,
   {Bhatt} B.~C.,    {Prabhu} T.~P.,  2008, Bulletin of the Astronomical
  Society of India, 36, 111

\bibitem[\protect\citeauthoryear{{Stephens}}{{Stephens}}{1974}]{Stephens1974}
{Stephens} M.~A.,  1974, Journal of the American Statistical Association, 69,
  730

\bibitem[\protect\citeauthoryear{{Szczygiel} \& {Fabrycky}}{{Szczygiel} \&
  {Fabrycky}}{2008}]{Szczygiel2008}
{Szczygiel} D.~M.,  {Fabrycky} D.~C.,  2008, VizieR Online Data Catalog, 837,
  71263

\bibitem[\protect\citeauthoryear{{Tamuz}, {Mazeh} \& {Zucker}}{{Tamuz}
  et~al.}{2005}]{Tamuz2005}
{Tamuz} O.,  {Mazeh} T.,    {Zucker} S.,  2005, MNRAS, 356, 1466

\bibitem[\protect\citeauthoryear{{Young}, {Genet}, {Boyd}, {Borucki},
  {Lockwood}, {Henry}, {Hall}, {Smith}, {Baliumas}, {Donahue} \&
  {Epand}}{{Young} et~al.}{1991}]{Young1991}
{Young} A.~T.,  {Genet} R.~M.,  {Boyd} L.~J.,  {Borucki} W.~J.,  {Lockwood}
  G.~W.,  {Henry} G.~W.,  {Hall} D.~S.,  {Smith} D.~P.,  {Baliumas} S.~L.,
  {Donahue} R.,    {Epand} D.~H.,  1991, PASP, 103, 221

\bibitem[\protect\citeauthoryear{{Zhang}, {Bianco}, {Lehner}, {Coehlo}, {Wang},
  {Mondal}, {Alcock}, {Axelrod}, {Byun}, {Chen}, {Cook}, {Dave}, {de Pater},
  {Porrata}, {Kim} \& et al.}{{Zhang} et~al.}{2008}]{Zhang2008ApJL}
{Zhang} Z.-W.,  {Bianco} F.~B.,  {Lehner} M.~J.,  {Coehlo} N.~K.,  {Wang}
  J.-H.,  {Mondal} S.,  {Alcock} C.,  {Axelrod} T.,  {Byun} Y.-I.,  {Chen}
  W.~P.,  {Cook} K.~H.,  {Dave} R.,  {de Pater} I.,  {Porrata} R.,  {Kim}
  D.-W.,    et al. 2008, ApJL, 685, L157

\bibitem[\protect\citeauthoryear{{Zhang}, {Kim}, {Wang}, {Lehner}, {Chen},
  {Byun}, {Alcock}, {Axelrod}, {Bianco}, {Cook}, {King}, {Lee}, {Lin},
  {Marshall}, {Schwamb}, {Wang} \& {Wen}}{{Zhang} et~al.}{2009}]{Zhang2009}
{Zhang} Z.-W.,  {Kim} D.-W.,  {Wang} J.-H.,  {Lehner} M.~J.,  {Chen} W.~P.,
  {Byun} Y.-I.,  {Alcock} C.~R.,  {Axelrod} T.,  {Bianco} F.~B.,  {Cook} K.~H.,
   {King} S.-K.,  {Lee} T.,  {Lin} H.-C.,  {Marshall} S.~L.,  {Schwamb} M.~E.,
  {Wang} S.-Y.,    {Wen} C.-Y.,  2009, In preparation

\end{thebibliography}
\label{lastpage}

\end{document}